\documentclass{article}

\usepackage{microtype}
\usepackage{graphicx}
\usepackage{subcaption}
\usepackage{booktabs} 

\usepackage{hyperref}
\usepackage{multirow}

\newcommand{\modelname}{BrainRVQ}

\usepackage[preprint]{icml2026}

\usepackage{amsmath}
\usepackage{amssymb}
\usepackage{mathtools}
\usepackage{amsthm}

\usepackage[capitalize,noabbrev]{cleveref}

\theoremstyle{plain}

\theoremstyle{definition}

\theoremstyle{remark}

\usepackage[textsize=tiny]{todonotes}

\icmltitlerunning{}

\begin{document}
\twocolumn[
  \icmltitle{BrainRVQ: A High-Fidelity EEG Foundation Model via Dual-Domain Residual Quantization and Hierarchical Autoregression}
  

  
  


  \icmlsetsymbol{equal}{*}

  \begin{icmlauthorlist}
    \icmlauthor{Mingzhe Cui}{yyy}
    \icmlauthor{Tao Chen}{equal,yyy,comp}
    \icmlauthor{Yang Jiao}{sch}
    \icmlauthor{Yiqin Wang}{yyy}
    \icmlauthor{Lei Xie}{yyy}
    \icmlauthor{Yi Pan}{sch}
    \icmlauthor{Luca Mainardi}{comp}
  \end{icmlauthorlist}

  \icmlaffiliation{yyy}{State Key Laboratory of Industrial Control Technology, Zhejiang University, Hangzhou, China}
  \icmlaffiliation{comp}{Department of Electronics, Information and Bioengineering, Politecnico di Milano, Milan, Italy}
  \icmlaffiliation{sch}{Shenzhen Institute of Advanced Technology, Chinese Academy of Sciences, Shenzhen, China}

  \icmlcorrespondingauthor{Tao Chen}{chentao98@zju.edu.cn}

  \icmlkeywords{Machine Learning, ICML}

  \vskip 0.3in
]



\printAffiliationsAndNotice{}  

\begin{abstract} 
Developing foundation models for electroencephalography (EEG) remains challenging due to the signal's low signal-to-noise ratio and complex spectro-temporal non-stationarity. Existing approaches often overlook the hierarchical latent structure inherent in neural dynamics, leading to suboptimal reconstruction of fine-grained information. In this work, we propose BrainRVQ, a general-purpose EEG foundation model pre-trained on a large-scale corpus of clinical EEG data. Unlike standard masked modeling, BrainRVQ features a Dual-Domain Residual Vector Quantization (DD-RVQ) tokenizer that disentangles temporal waveforms and spectral patterns into hierarchical discrete codes. We further introduce a hierarchical autoregressive pre-training objective that learns to reconstruct these codes in a coarse-to-fine manner, utilizing an importance-guided curriculum masking strategy to prioritize information-rich neural events over background noise. Extensive experiments across 8 diverse downstream datasets demonstrate that BrainRVQ consistently outperforms state-of-the-art baselines, validating its effectiveness in learning robust and generalizable neural representations. Our code and model weights are available:
https://github.com/keqicmz/BrainRVQ
\end{abstract}

\section{Introduction}
Electroencephalography (EEG) provides a non-invasive interface for monitoring millisecond-level neural dynamics~\cite{niedermeyer2005electroencephalography}. This high temporal resolution has fueled advancements across diverse domains, ranging from seizure detection~\cite{shoeb2009application}, sleep staging~\cite{aboalayon2016sleep}, emotion recognition~\cite{zheng2015investigating} and motor imagery classification~\cite{pfurtscheller2001motor}. However, decoding EEG signals remains a formidable challenge for machine learning due to their inherently low signal-to-noise ratios (SNR), complex non-stationarity, and substantial variability across subjects~\cite{lotte2007review}. These characteristics, combined with the scarcity of large-scale labeled data, have historically constrained the development of generalizable EEG decoding models in neuroscience.

To address the labeled data bottleneck, self-supervised learning (SSL) has emerged as a promising paradigm. Early discriminative approaches, such as BENDR~\cite{kostas2021bendr}, employed contrastive learning to extract transferable features from unlabeled recordings. More recently, generative approaches inspired by Masked Image Modeling (MIM) and Large Language Models (LLMs) have gained prominence. Pioneering works like LaBraM~\cite{jiang2024large} introduced neural tokenizers to convert continuous EEG into discrete tokens, enabling BERT-style pre-training. BrainBERT~\cite{wang2023brainbert} and Brant~\cite{zhang2023brant} further explored distinct masking strategies and architectural designs to capture temporal dependencies. Subsequent efforts have focused on scaling and architectural innovations: REVE~\cite{ouahidi2025reve} scaled pre-training to over 25,000 subjects, CBraMod~\cite{wang2024cbramod} proposed criss-cross attention to separately model spatial and temporal dependencies.

Despite these advances, existing EEG foundation models face critical limitations in representation fidelity. First, most current approaches rely on single-domain tokenization, processing signals either strictly in the time domain or the frequency domain. While time-domain tokenizers excel at capturing transient events like epileptic spikes, they often struggle to represent global spectral patterns. Conversely, frequency-domain methods capture oscillatory rhythms but sacrifice temporal precision. Consequently, this single-perspective quantization fails to capture the intricate spectro-temporal coupling of neural dynamics, leading to significant information loss and suboptimal reconstruction of complex brain signals.

Second, the discretization capacity of existing models remains limited. Standard neural tokenizers typically employ single-layer vector quantization to project continuous signals into discrete latent codes. This flat quantization lacks the capacity to encode the high-dimensional variability of EEG signals. Drawing inspiration from high-fidelity audio generation, Residual Vector Quantization (RVQ) offers a robust solution by employing a cascade of codebooks to approximate signals with increasing precision. However, directly applying RVQ to EEG pre-training is non-trivial. EEG signals exhibit an inherent hierarchy characterized by dominant rhythms and subtle details, yet are significantly contaminated by background noise. Standard masked modeling objectives treat all residual codes independently and equally, which fails to capture the coarse-to-fine semantic dependency and often results in the model wasting capacity on reconstructing background noise rather than meaningful neural events.

In this work, we propose BrainRVQ, a high-fidelity EEG foundation model designed to resolve these challenges through Dual-Domain Residual Quantization and Hierarchical Autoregression. To overcome the information loss caused by single-domain processing, we introduce a Dual-Domain Residual Vector Quantization (DD-RVQ) tokenizer. This module disentangles EEG signals into hierarchical discrete codes across both time and frequency domains simultaneously, ensuring that both fine-grained temporal transients and global spectral oscillations are preserved. Furthermore, to address the limitations of flat quantization and independent prediction, we propose a Hierarchical Autoregressive Pre-training objective. Instead of predicting tokens independently, our model learns to reconstruct RVQ codes in a coarse-to-fine manner using teacher forcing, explicitly modeling the dependency between layers. This is coupled with an importance-guided curriculum masking strategy, which dynamically prioritizes information-rich regions, enabling the model to learn robust representations from a cleaner neural manifold.

The main contributions of this paper are summarized as follows: \begin{itemize} \item \textbf{Dual-Domain Residual Vector Quantization (DD-RVQ):} We propose a novel tokenizer that performs hierarchical quantization in both time and frequency domains. This dual-perspective design mitigates the information loss inherent in single-domain approaches, enabling high-fidelity encoding of spectro-temporal dynamics. \item \textbf{Hierarchical Autoregressive Pre-training:} We introduce a generative pre-training objective that models the coarse-to-fine dependency of residual codes. By utilizing teacher forcing to predict finer-grained codes conditioned on coarser priors, the model captures the structural hierarchy of neural signals more effectively than standard independent masking. \item \textbf{Importance-Guided Curriculum Masking:} We propose an adaptive masking strategy that prioritizes high-information patches based on spectral neural content and temporal signal complexity. This curriculum-based approach effectively suppresses background noise, forcing the model to focus on semantically meaningful neural events. \item \textbf{Comprehensive Validation:} We pre-train BrainRVQ on a large-scale corpus of clinical EEG data. Extensive experiments across 8 diverse downstream datasets—including seizure detection, emotion recognition, and sleep staging—validate the superior generalization capabilities and advanced performance of our model compared to state-of-the-art baselines. \end{itemize}

\section{Methodology}
\subsection{Overview}

\begin{figure*}[t]
  \centering
  \includegraphics[width=0.9\textwidth]{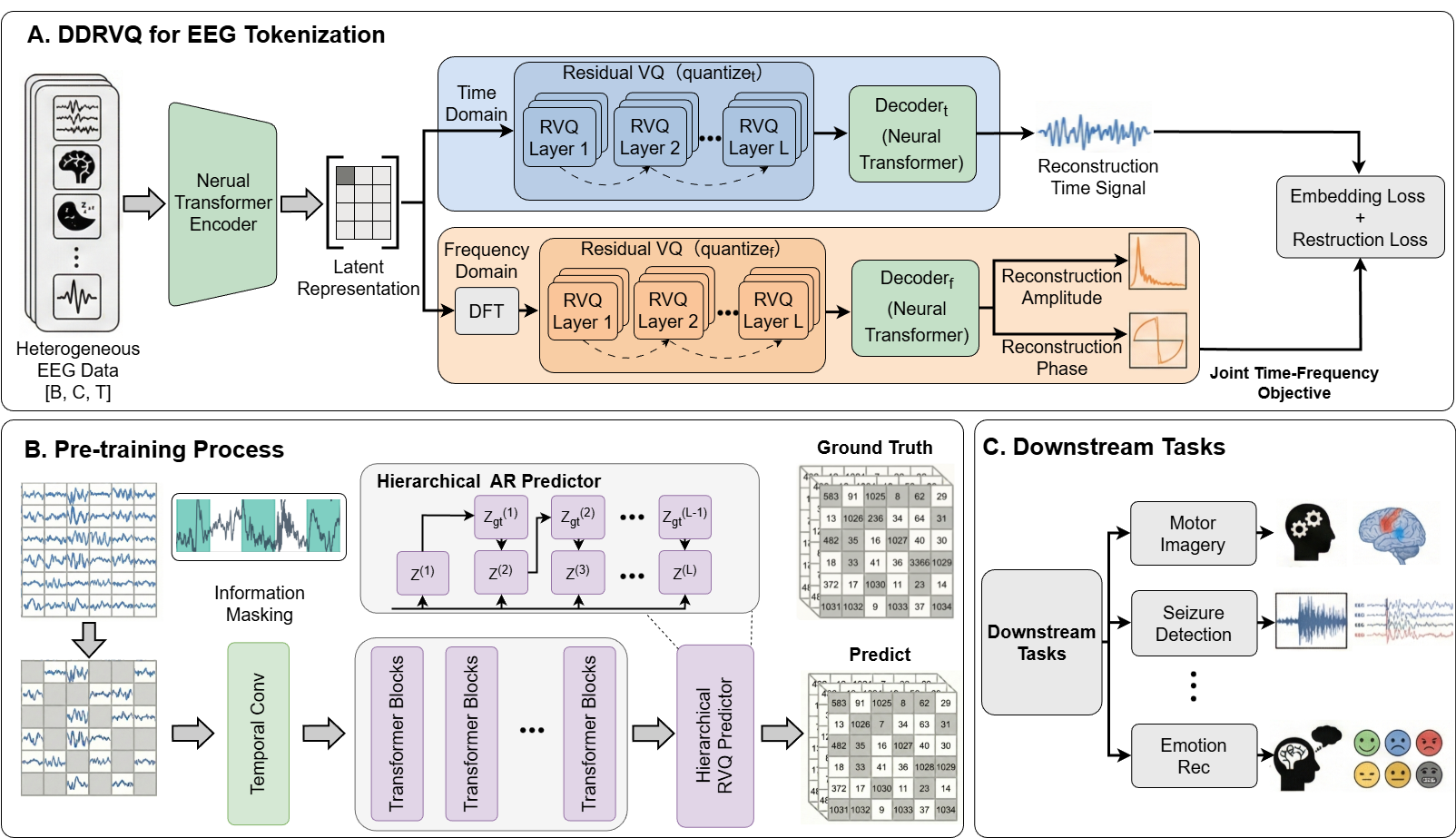}
  \caption{
    \textbf{The overall architecture of the BrainRVQ framework.}
    \textbf{(A) DDRVQ for EEG Tokenization:} The module employs a DDRVQ mechanism to discretize EEG signals. It extracts features simultaneously in time and frequency domains and is optimized via a joint objective of reconstruction and embedding loss.
    \textbf{(B) Pre-training Stage:} We introduce a Hierarchical Autoregressive Masked Modeling objective. The model learns to predict residual tokens in a coarse-to-fine manner, guided by an Importance-Guided Curriculum Masking strategy that prioritizes high-information regions.
    \textbf{(C) Downstream Adaptation:} The pre-trained encoder serves as a general-purpose feature extractor.
  }
  \label{fig:framework} 
\end{figure*}

The overall framework of BrainRVQ is illustrated in \cref{fig:framework}. Our approach consists of three integrated stages designed to learn high-fidelity representations from continuous EEG signals. First, input signals are processed by the DD-RVQ tokenizer, which discretizes neural dynamics into hierarchical codes across both temporal and spectral domains. Second, we employ an importance-guided curriculum masking strategy to dynamically select information-rich patches based on spectral neural content and temporal complexity. Finally, the masked representations are fed into a Transformer encoder optimized via a hierarchical autoregressive objective, which reconstructs the discrete codes in a coarse-to-fine manner using teacher forcing. The following subsections detail each component.
\subsection{EEG Signal Patching and Preprocessing}

Given a raw EEG recording $\mathbf{X} \in \mathbb{R}^{C \times T}$, where $C$ denotes the number of channels and $T$ denotes the number of temporal samples, we segment each channel into non-overlapping patches to enable local feature extraction. Specifically, we apply a sliding window of length $P$ without overlap to partition each channel into patches, yielding:
\begin{equation}
\boldsymbol{x} = \{x_{c,a} \in \mathbb{R}^{P} \mid c = 1, \ldots, C,\ a = 1, \ldots, A\},
\end{equation}
where $A = \lfloor \frac{T}{P} \rfloor$ represents the number of patches per channel, and the total number of patches is $|\boldsymbol{x}| = C \times A$. In our implementation, we set $P = 200$ corresponding to 1-second windows at 200\,Hz sampling rate. The patched signal is reshaped into a 3D tensor $\mathbf{X}_p \in \mathbb{R}^{C \times A \times P}$, where each element $x_{c,a}$ represents the $a$-th temporal segment of channel $c$. 

\subsection{Dual-Domain Residual Vector Quantization}

To construct a semantically rich codebook for EEG signals, we propose a DD-RVQ mechanism that addresses two fundamental limitations of standard VQ-VAE when applied to neural signals: (1) \textbf{Domain Incompleteness}: processing signals solely in the temporal domain fails to capture frequency-specific features such as alpha rhythms (8--12\,Hz) or beta oscillations (13--30\,Hz), which are critical for cognitive state recognition; (2) \textbf{Quantization Bottleneck}: single-layer VQ imposes a hard capacity ceiling, often leading to codebook collapse where only a small subset of codes are utilized, resulting in severe information loss. Our DD-RVQ addresses these challenges through three synergistic components: a shared encoder that extracts joint time-frequency embeddings, a residual quantization hierarchy with separate temporal and frequency codebooks, and domain-specific decoders with complementary reconstruction objectives.

\subsubsection{Dual-Domain Encoding}
EEG signals exhibit fundamentally distinct characteristics across domains. The temporal waveform encodes transient event-related potentials (ERPs) and rapid state transitions, while the frequency spectrum reveals sustained oscillatory patterns tied to cognitive processes. To leverage both representations, we employ a shared-encoder, dual-branch architecture.

Given a normalized patch $\bar{x}_{c,a} \in \mathbb{R}^{P}$, we first apply the Discrete Fourier Transform (DFT) to obtain its spectral representation:
\begin{equation}
\mathbf{X}_{\text{freq}}[k] = \sum_{n=0}^{P-1} \bar{x}_{c,a}[n] \cdot e^{-j\frac{2\pi kn}{P}}, \quad k = 0, 1, \ldots, P-1,
\end{equation}
where $j$ denotes the imaginary unit. From the complex-valued spectrum, we extract the amplitude $A[k] = |\mathbf{X}_{\text{freq}}[k]|$ and phase $\phi[k] = \arg(\mathbf{X}_{\text{freq}}[k])$ as reconstruction targets. The temporal patch $\bar{x}_{c,a}$ is processed through a shared Transformer encoder, yielding a unified latent representation $\mathbf{h} \in \mathbb{R}^{d}$. This representation is then passed through two parallel Residual Vector Quantizers—a temporal quantizer $\mathcal{Q}_t$ and a frequency quantizer $\mathcal{Q}_f$—producing dual discrete token sequences. The quantized representations are subsequently decoded by domain-specific decoders to reconstruct the temporal waveform and the frequency spectra, respectively. This shared-encoder, dual-branch design enables learning unified representations that capture both transient dynamics and oscillatory structure.

\subsubsection{Residual Vector Quantization: Hierarchical Coarse-to-Fine Discretization}

Standard single-layer VQ maps each continuous embedding to a single discrete code, limiting representational capacity. To overcome this bottleneck, we employ RVQ, which iteratively refines the representation through a hierarchy of codebooks $\{\mathbf{V}^{(l)}\}_{l=1}^{L}$, where each codebook $\mathbf{V}^{(l)} = \{v_k^{(l)} \in \mathbb{R}^{d}\}_{k=1}^{K_l}$ contains $K_l$ code vectors at level $l$. The quantization proceeds recursively:
\begin{equation}
\begin{aligned}
z_i^{(l)} &= \arg\min_{k \in \{1, \ldots, K_l\}} \|\mathbf{r}^{(l-1)} - \mathbf{v}_k^{(l)}\|_2, \\
\mathbf{r}^{(l)} &= \mathbf{r}^{(l-1)} - \mathbf{v}_{z_i^{(l)}}^{(l)}, \quad \text{with } \mathbf{r}^{(0)} = \tilde{\mathbf{e}},
\end{aligned}
\end{equation}
where $\mathbf{r}^{(l)}$ denotes the residual after quantization at level $l$, and $z_i^{(l)} \in \{1, \ldots, K_l\}$ is the selected code index. The final quantized representation is the sum of all selected codes:
\begin{equation}
\mathbf{h}_q = \sum_{l=1}^{L} \mathbf{v}_{z_i^{(l)}}^{(l)}.
\end{equation}
This hierarchical structure enables coarse-to-fine modeling: the first layer captures global patterns, while subsequent layers refine local details. Critically, we maintain \textit{separate} RVQ branches for temporal and frequency domains, denoted as $\mathbf{h}_q^{(t)}$ and $\mathbf{h}_q^{(f)}$, each producing $L$ token indices per patch: $\mathbf{z}^{(t)} = [z^{(1)}, \ldots, z^{(L)}]$ and $\mathbf{z}^{(f)} = [z^{(1)}, \ldots, z^{(L)}]$.

\subsubsection{Reconstruction}

To supervise the learning of meaningful codebooks, we employ three complementary reconstruction objectives that enforce consistency across temporal and spectral views. The temporal decoder reconstructs the normalized waveform from $\mathbf{h}_q^{(t)}$:
\begin{equation}
\mathcal{L}_{\text{time}} = \|\text{Decoder}_t(\mathbf{h}_q^{(t)}) - \bar{x}_{c,a}\|_2^2,
\end{equation}
encouraging the temporal codebook to preserve the original signal morphology. The frequency decoder reconstructs both amplitude and phase spectra from $\mathbf{h}_q^{(f)}$:
\begin{equation}
\mathcal{L}_{\text{freq}} = \|y_A - A\|_2^2 + \|y_\phi - \phi\|_2^2,
\end{equation}
where $y_A = \text{Decoder}_f^{(A)}(\mathbf{h}_q^{(f)})$ and $y_\phi = \text{Decoder}_f^{(\phi)}(\mathbf{h}_q^{(f)})$ are the predicted amplitude and phase. This dual reconstruction ensures the frequency codebook captures spectral energy distribution and phase relationships critical for oscillatory patterns. 

The total training objective integrates these reconstruction terms with a commitment loss to regularize the discrete latent space. Given the hierarchical nature of our quantization, we enforce the commitment constraint at every residual level $l$. The total loss function is formulated as:
\begin{equation}
\mathcal{L}_{\text{total}} = \mathcal{L}_{\text{time}} + \mathcal{L}_{\text{freq}} + \beta \sum_{d \in \{t, f\}} \sum_{l=1}^{L} \|\text{sg}[\mathbf{r}_d^{(l-1)}] - \mathbf{v}_d^{(l)}\|_2^2,
\end{equation}
where $\text{sg}[\cdot]$ denotes the stop-gradient operator used to prevent encoder collapse. The summation term aggregates the quantization errors across all $L$ layers for both temporal and frequency domains, where $\mathbf{r}_d^{(l-1)}$ is the residual input to layer $l$ and $\mathbf{v}_d^{(l)}$ is the selected code vector. By enforcing consistency across temporal waveforms, amplitude spectra, and phase spectra, this multi-view constraint ensures that the learned tokens encapsulate rich, noise-robust representations of neural dynamics, enabling effective downstream transfer.

\subsection{Hierarchical Autoregressive Pre-training}

Standard masked autoencoding predicts each token independently from the same contextualized embedding, discarding the coarse-to-fine structure inherent in RVQ representations. The first-layer code $z_i^{(1)}$ captures global patterns, which should inform predictions of subsequent codes to refine local details. To exploit this hierarchy, we design a pre-training framework that models inter-layer dependencies via autoregressive factorization.

\subsubsection{Autoregressive Factorization}

Given the set of masked patch indices $\mathcal{M}$ and visible context $\mathbf{X}_{\setminus\mathcal{M}}$, standard approaches formulate the objective as:
\begin{equation}
\mathcal{L}_{\text{MAE}} = \sum_{i \in \mathcal{M}} \sum_{l=1}^{L} -\log P(z_i^{(l)} | \mathbf{X}_{\setminus\mathcal{M}}),
\end{equation}
where each layer's token $z_i^{(l)}$ is predicted independently. This independence assumption ignores the hierarchical prior established during tokenization. To address this, we factorize the joint distribution of all tokens at position $i$ as:
\begin{equation}
P(\mathbf{z}_i | \mathbf{X}_{\setminus\mathcal{M}}) = \prod_{l=1}^{L} P(z_i^{(l)} | z_i^{(1:l-1)}, \mathbf{X}_{\setminus\mathcal{M}}),
\end{equation}
where $z_i^{(1:l-1)} = [z_i^{(1)}, \ldots, z_i^{(l-1)}]$ denotes codes from preceding layers. This formulation ensures that each layer's prediction is conditioned on coarser priors, enabling progressive refinement from global structure to local detail.

Architecturally, we employ a shared Transformer encoder $f_\theta$ followed by $L$ layer-specific prediction heads $\{g_\theta^{(l)}\}_{l=1}^{L}$. To predict the token at layer $l$, we augment the encoder output with the \textit{cumulative} embeddings from all preceding layers:
\begin{equation}
\hat{z}_i^{(l)} = g_\theta^{(l)}\left( \text{LN}^{(l)}\left( \mathbf{h}_i + \sum_{k=1}^{l-1} \text{Embed}^{(k)}(z_i^{(k)}) \right) \right),
\end{equation}
where $\mathbf{h}_i = f_\theta(\mathbf{X}_{\setminus\mathcal{M}})_i$ is the encoder output for position $i$, $\text{Embed}^{(k)}(\cdot)$ denotes a learnable embedding that maps the discrete code index $z_i^{(k)}$ to the hidden dimension, and $\text{LN}^{(l)}$ is a layer-specific normalization. For the first layer ($l=1$), no conditioning is applied and the prediction head directly operates on $\mathbf{h}_i$. The cumulative sum encodes the partial reconstruction from all coarser levels, providing a progressively refined structural prior.

\subsubsection{Teacher-Forcing Optimization}

Direct autoregressive training would require sampling $\hat{z}_i^{(l-1)}$ to condition layer $l$, leading to error accumulation from noisy early-layer predictions. To stabilize optimization, we employ teacher forcing: during training, we condition on ground-truth codes $z_i^{(1:l-1)}$ obtained from the tokenizer rather than model predictions. The layer-wise loss becomes:
\begin{equation}
\mathcal{L}^{(l)} = \sum_{i \in \mathcal{M}} -\log P_\theta(z_i^{(l)} | z_i^{(1:l-1)}, \mathbf{X}_{\setminus\mathcal{M}}).
\end{equation}

The total pre-training objective aggregates losses across all $L$ layers with depth-dependent weights:
\begin{equation}
\mathcal{L}_{\text{HAR}} = \sum_{l=1}^{L} \lambda_l \cdot \mathcal{L}^{(l)},
\end{equation}
where $\lambda_l = 2^{-(l-1)}$ assigns higher importance to coarser layers, as errors in early predictions propagate through the autoregressive chain. At inference time, the model operates in a fully autoregressive manner: it first predicts $\hat{z}_i^{(1)}$, retrieves the corresponding embedding to condition layer 2, and iterates through all $L$ layers, faithfully replicating the generative factorization.

\subsubsection{Importance-Guided Curriculum Masking}

Standard masked language modeling employs uniform random masking, sampling patches with equal probability regardless of their information content. While this ensures unbiased coverage, it is suboptimal for physiological signals where informative neural events are sparse and temporally localized, including epileptic spikes, sleep spindles, and event-related potentials. Masking uninformative background yields trivial reconstruction tasks, whereas selectively masking high-information regions forces the encoder to reason about complex neural dynamics. To address this, we propose an importance-guided curriculum masking strategy that transitions from exploration to exploitation as training proceeds.

\paragraph{Physiology-Aware Importance Scoring.}
To quantify the information density of each patch, we design a composite scoring function integrating frequency-domain and time-domain characteristics. In the frequency domain, we compute the power spectral density via DFT and extract two metrics: (1) the neural band ratio, measuring power concentration within physiologically relevant bands; and (2) the artifact penalty, quantifying contamination from baseline drift and muscle artifacts. In the time domain, we employ Hjorth parameters, which provide computationally efficient estimates of mean frequency and spectral bandwidth, respectively. We additionally compute an irregularity metric capturing signal non-stationarity. The final importance score aggregates these metrics:
\begin{equation}
S_{c,a} = \sum_{m} \alpha_m \cdot \tilde{S}_m,
\end{equation}
where $\tilde{S}_m$ denotes the min-max normalized score for metric $m$, and $\{\alpha_m\}$ are weighting coefficients that prioritize neural band content and artifact-free segments.

\paragraph{Curriculum via Importance-Weighted Sampling.}
To implement curriculum learning, we adopt a probabilistic sampling strategy that balances information-guided selection with stochastic exploration. The combined score for each patch is:
\begin{equation}
\hat{S}_{c,a}(t) = w(t) \cdot S_{c,a} + (1 - w(t)) \cdot U_{c,a},
\end{equation}
where $U_{c,a} \sim \text{Uniform}(0, 1)$ introduces randomness, and $w(t)$ is a curriculum weight that increases linearly during training. Early in training, the sampling distribution is nearly uniform, encouraging broad exploration. As training progresses, the distribution becomes increasingly biased toward high-importance patches, forcing the model to focus on information-dense regions. This curriculum strategy improves both convergence speed and final representation quality by balancing global coverage with focused learning on challenging neural events.The detailed formulation of importance scoring metrics is provided in Appendix~\ref{app:method_details}.

\section{Experiments}

\subsection{Pre-training}

\textbf{Dataset.}
We pre-train our model on the large-scale Temple University Hospital EEG Corpus (TUEG)~\cite{obeid2016temple} (v2.0.1). As one of the largest publicly available clinical EEG archives, it comprises 26,846 sessions collected from 14,987 unique patients, spanning a cumulative duration exceeding 27,000 hours. The corpus exhibits significant real-world heterogeneity, featuring diverse acquisition parameters with sampling rates ranging from 250 Hz to 1024 Hz and over 40 differing channel configurations. This substantial scale and inherent clinical variability provide a robust foundation for learning generalizable neural representations across diverse physiological and pathological states.

\textbf{Preprocessing.}
We follow standard preprocessing protocols to ensure data quality. Recordings shorter than 5 minutes are excluded, and the first and last minute of each session are discarded to remove boundary artifacts. We retain 19 channels conforming to the international 10-20 system (Fp1, Fp2, F7, F3, Fz, F4, F8, T3, C3, Cz, C4, T4, T5, P3, Pz, P4, T6, O1, O2). A band-pass filter (0.3--75 Hz) is applied to remove low-frequency drift and high-frequency noise, followed by a notch filter at 60 Hz to suppress power line interference. All signals are resampled to 200 Hz and segmented into 30-second non-overlapping samples. To further ensure quality, samples containing any data point with absolute amplitude exceeding 100 $\mu$V are discarded. Finally, amplitudes are normalized by dividing by 100 $\mu$V. After preprocessing, 1,109,545 samples (approximately 9,200 hours) are retained for pre-training.

\textbf{Implementation Details.}
We implemented BrainRVQ using Python 3.10 and PyTorch 2.1.2 + CUDA 11.8. All experiments are conducted on NVIDIA A800 (80GB) GPUs. For the DD-RVQ Tokenizer, we use a 12-layer Transformer encoder with embedding dimension 200 and a 3-layer Transformer decoder. Both temporal and frequency codebooks employ 3-layer RVQ with vocabulary sizes 8192 and embedding dimension 64. The DDRVQ is trained for 20 epochs with batch size 128, taking approximately 12 hours on six NVIDIA A800 GPUs. For pre-training, we use a 12-layer Transformer encoder with hidden dimension 200 and 10 attention heads. Pre-training runs for 20 epochs with batch size 64, taking approximately 20 hours on six NVIDIA A800 GPUs. More details can be found in Appendix~\ref{app:pretrain_analysis}.

\subsection{Experimental Setup}

\textbf{Downstream Tasks.}
We evaluate BrainRVQ on a comprehensive suite of 8 downstream datasets spanning diverse BCI application domains, as summarized in Table~\ref{tab:datasets}. For neurological diagnosis, we include TUAB~\citep{obeid2016temple}, TUEV~\citep{obeid2016temple}, and CHB-MIT~\citep{shoeb2009application}. For affective computing, we adopt the 5-class emotion recognition benchmark SEED-V~\citep{liu2021comparing}. For motor imagery decoding, we evaluate on PhysioNet~\citep{schalk2004bci2000}, SHU-MI~\citep{malarge}, and BCICIV-2a~\citep{tangermann2012review}, covering various electrode configurations and classification scenarios. Finally, for cognitive state monitoring, we include the Mental Workload dataset~\citep{zyma2019electroencephalograms}. This diverse protocol rigorously tests the generalization capability across different recording setups and paradigms. Detailed configurations are provided in Appendix~\ref{app:downstream}.

\textbf{Baselines.}
We compare BrainRVQ against two categories of methods. The first includes supervised baselines trained from scratch: EEGNet~\citep{lawhern2018eegnet} and ST-Transformer~\citep{song2021transformer}. The second comprises self-supervised foundation models pre-trained on large-scale EEG corpora: BENDR~\citep{kostas2021bendr}, BIOT~\citep{yang2023biot}, LaBraM~\citep{jiang2024large}, and CBraMod~\citep{wang2024cbramod}. All foundation models are pre-trained on the TUH EEG corpus and fine-tuned following their official protocols to ensure fair comparison. Implementation details are in Appendix~\ref{app:downstream}.

\begin{table}[t]
  \caption{Overview of the 8 downstream datasets across diverse BCI tasks.}
  \label{tab:datasets}
  \begin{center}
    \begin{small}
        \resizebox{\columnwidth}{!}{
        \begin{tabular}{l@{\hskip 1.5em}l@{}cccc}
          \toprule
          Task & Dataset & \#Ch & Win(s) & \#Samps & Type \\
          \midrule
          Abnormal Det. & 
            \begin{tabular}[t]{@{}l@{}} TUAB \\ \upshape \citep{obeid2016temple} \end{tabular} 
            & 16 & 10 & 409,455 & Binary \\
          \noalign{\vskip 3pt}
          Event Class.  & 
            \begin{tabular}[t]{@{}l@{}} TUEV \\ \upshape \citep{obeid2016temple} \end{tabular} 
            & 16 & 5  & 112,491 & 6-Class \\
          \noalign{\vskip 3pt}
          Seizure Det.  & 
            \begin{tabular}[t]{@{}l@{}} CHB-MIT \\ \upshape \citep{shoeb2009application} \end{tabular}  
            & 16 & 10 & 326,993 & Binary \\
          \noalign{\vskip 3pt}
          Emotion Rec.  & 
            \begin{tabular}[t]{@{}l@{}} SEED-V \\ \upshape \citep{liu2021comparing} \end{tabular} 
            & 62 & 1  & 117,744 & 5-Class \\

          \multirow{3}{*}{Motor Imagery}
          
                                         & 
            \begin{tabular}[t]{@{}l@{}} PhysioNet \\ \upshape \citep{schalk2004bci2000} \end{tabular}
            & 64 & 4  & 9,837   & 4-Class \\
          \noalign{\vskip 3pt}
                                         & 
            \begin{tabular}[t]{@{}l@{}} SHU-MI \\ \upshape \citep{malarge} \end{tabular}       
            & 32 & 4  & 11,988  & Binary \\
          \noalign{\vskip 3pt}
                                         & 
            \begin{tabular}[t]{@{}l@{}} BCICIV-2a \\ \upshape \citep{tangermann2012review} \end{tabular}  
            & 22 & 4  & 5,088   & 4-Class \\

          Workload      & 
            \begin{tabular}[t]{@{}l@{}} Mental \\ \upshape \citep{zyma2019electroencephalograms} \end{tabular}    
            & 20 & 5  & 1,707   & Binary \\
          \bottomrule
        \end{tabular}
        }
    \end{small}
  \end{center}
  \vskip -0.1in
\end{table}
\textbf{Implementation and Metrics.}
We adopt Balanced Accuracy, AUC-PR, and AUROC for binary classification, 
and Balanced Accuracy, Cohen's Kappa, and Weighted F1 for multi-class classification. 
AUROC and Kappa serve as the monitor scores for model selection, respectively. 
All results are reported as mean $\pm$ standard deviation over five random seeds.

\subsection{Results}

\begin{table*}[t]
  \caption{Main evaluation results on four representative downstream tasks. The table is divided into binary classification (top) and multi-class classification (bottom). Best results are in \textbf{bold}, second best are \underline{underlined}.}
  \label{tab:main_results}
  \centering
  \resizebox{\textwidth}{!}{
  \begin{tabular}{l|ccc|ccc}
    \toprule
    \multirow{2}{*}{\textbf{Method}} & \multicolumn{3}{c|}{\textbf{Mental Workload}} & \multicolumn{3}{c}{\textbf{CHB-MIT}} \\
    \cmidrule(lr){2-4} \cmidrule(lr){5-7} 
    & Bal. Acc & AUC-PR & AUROC & Bal. Acc & AUC-PR & AUROC \\
    \midrule
    EEGNet & 0.677 $\pm$ 0.012 & 0.576 $\pm$ 0.010 & 0.732 $\pm$ 0.011 & 0.566 $\pm$ 0.011 & 0.191 $\pm$ 0.018 & 0.805 $\pm$ 0.014 \\
    ST-Transformer & 0.663 $\pm$ 0.017 & 0.567 $\pm$ 0.026 & 0.713 $\pm$ 0.017 & 0.592 $\pm$ 0.020 & 0.142 $\pm$ 0.009 & 0.824 $\pm$ 0.049 \\
    BENDR & 0.568 $\pm$ 0.045 & 0.366 $\pm$ 0.067 & 0.568 $\pm$ 0.045 & 0.561 $\pm$ 0.043 & 0.307 $\pm$ 0.124 & 0.863 $\pm$ 0.053 \\
    BIOT & 0.688 $\pm$ 0.019 & 0.600 $\pm$ 0.020 & 0.754 $\pm$ 0.014 & 0.707 $\pm$ 0.046 & 0.328 $\pm$ 0.046 & 0.876 $\pm$ 0.028 \\
    LaBraM & 0.691 $\pm$ 0.013 & 0.600 $\pm$ 0.016 & 0.772 $\pm$ 0.009 & 0.708 $\pm$ 0.036 & 0.329 $\pm$ 0.040 & 0.868 $\pm$ 0.020 \\
    CBraMod & \underline{0.726 $\pm$ 0.013} & \underline{0.627 $\pm$ 0.010} & \underline{0.791 $\pm$ 0.007} & \textbf{0.740 $\pm$ 0.028} & \underline{0.369 $\pm$ 0.038} & \underline{0.889 $\pm$ 0.015} \\
    \textbf{\modelname{} (Ours)} & \textbf{0.747 $\pm$ 0.011} & \textbf{0.758 $\pm$ 0.012} & \textbf{0.862 $\pm$ 0.010} & \underline{0.709 $\pm$ 0.040} & \textbf{0.465 $\pm$ 0.036} & \textbf{0.928 $\pm$ 0.024} \\
    
    \midrule
    
    \multirow{2}{*}{\textbf{Method}} & \multicolumn{3}{c|}{\textbf{TUEV}} & \multicolumn{3}{c}{\textbf{BCICIV-2a}} \\
    \cmidrule(lr){2-4} \cmidrule(lr){5-7}
    & Bal. Acc & Kappa & W-F1 & Bal. Acc & Kappa & W-F1 \\
    \midrule
    EEGNet & 0.388 $\pm$ 0.014 & 0.358 $\pm$ 0.016 & 0.654 $\pm$ 0.012 & 0.448 $\pm$ 0.009 & 0.269 $\pm$ 0.012 & 0.423 $\pm$ 0.011 \\
    ST-Transformer & 0.398 $\pm$ 0.023 & 0.377 $\pm$ 0.031 & 0.682 $\pm$ 0.019 & 0.458 $\pm$ 0.015 & 0.273 $\pm$ 0.020 & 0.447 $\pm$ 0.014 \\
    BENDR & 0.436 $\pm$ 0.025 & 0.427 $\pm$ 0.024 & 0.676 $\pm$ 0.022 & 0.490 $\pm$ 0.007 & 0.320 $\pm$ 0.009 & 0.484 $\pm$ 0.007 \\
    BIOT & 0.528 $\pm$ 0.023 & 0.527 $\pm$ 0.025 & 0.749 $\pm$ 0.008 & 0.475 $\pm$ 0.009 & 0.300 $\pm$ 0.014 & 0.461 $\pm$ 0.013 \\
    LaBraM & 0.641 $\pm$ 0.007 & 0.664 $\pm$ 0.009 & 0.831 $\pm$ 0.005 & 0.487 $\pm$ 0.009 & 0.316 $\pm$ 0.015 & 0.476 $\pm$ 0.010 \\
    CBraMod & \underline{0.667 $\pm$ 0.011} & \underline{0.677 $\pm$ 0.010} & \underline{0.834 $\pm$ 0.006} & \underline{0.514 $\pm$ 0.007} & \underline{0.352 $\pm$ 0.009} & \underline{0.498 $\pm$ 0.009} \\
    \textbf{\modelname{} (Ours)} & \textbf{0.668 $\pm$ 0.015} & \textbf{0.690 $\pm$ 0.008} & \textbf{0.840 $\pm$ 0.005} & \textbf{0.541 $\pm$ 0.008} & \textbf{0.388 $\pm$ 0.008} & \textbf{0.533 $\pm$ 0.012} \\
    \bottomrule
  \end{tabular}
  }
\end{table*}

To comprehensively validate the capability and generalizability of BrainRVQ, we evaluate it against baselines across a diverse suite of eight publicly available BCI datasets. In this section, we focus on four representative downstream tasks, listed in Table~\ref{tab:main_results}. In all experiments, we ensure strict consistency in the splits of training, validation, and test sets for every method. The comprehensive results on the additional datasets are detailed in Appendix~\ref{app:downstream}.

\textbf{Mental Workload Detection.}
We use the Mental Workload dataset for cognitive state assessment. BrainRVQ achieves state-of-the-art performance on this task. Specifically, BrainRVQ obtains a significant performance improvement compared to the best baseline CBraMod (0.862 vs. 0.791 in AUROC and 0.758 vs. 0.627 in AUC-PR). The substantial gain in AUC-PR demonstrates that our model excels at distinguishing positive samples from negative ones across all decision thresholds, which is critical for practical deployment in human-machine interaction systems.

\textbf{Seizure Detection.}
CHB-MIT is a highly imbalanced clinical dataset for pediatric seizure detection. BrainRVQ achieves the best performance on ranking metrics, obtaining 0.928 in AUROC and 0.465 in AUC-PR, compared to the best baseline CBraMod (0.889 in AUROC and 0.369 in AUC-PR). The remarkable 26.0\% relative improvement in AUC-PR indicates that our importance-guided curriculum masking effectively forces the model to attend to sparse but discriminative ictal patterns rather than the dominant interictal background.

\textbf{Event Type Classification.}
TUEV is a challenging 6-class clinical event classification task from the Temple University Hospital corpus. BrainRVQ achieves the best performance on all metrics. Specifically, BrainRVQ obtains a Cohen's Kappa of 0.690, surpassing the best baseline CBraMod (0.677 in Kappa). This improvement demonstrates the effectiveness of our hierarchical autoregressive pre-training in capturing diverse clinical EEG patterns including spike-and-wave complexes, generalized periodic discharges, and artifact signatures.

\textbf{Motor Imagery Classification.}
We use BCICIV-2a for evaluation on 4-class motor imagery classification. BrainRVQ achieves a significant performance gain compared to the best baseline CBraMod (0.388 vs. 0.352 in Cohen's Kappa and 0.533 vs. 0.498 in Weighted F1). This 10.2\% relative improvement in Kappa demonstrates that our Dual-Domain RVQ effectively captures both the transient event-related desynchronization and the sustained sensorimotor rhythms that characterize motor intention, which single-domain tokenizers tend to conflate.

\subsection{Ablation Study}
To validate the efficacy of each component in our proposed framework, we conducted comprehensive ablation studies across four diverse datasets: Mental Workload, CHB-MIT, TUEV, and BCICIV-2a. The quantitative results are summarized in Table~\ref{tab:ablation}, with corresponding visualizations provided in Figure~\ref{fig:ablation}. More comprehensive ablation results are provided in Appendix~\ref{app:ablation} and Appendix~\ref{app:rvq_depth}.

\begin{table}[t]
  \caption{Ablation study results. AUROC for binary and Cohen's Kappa for multi-class tasks.}
  \label{tab:ablation}
  \begin{center}
    \begin{small}
        \resizebox{\columnwidth}{!}{
          \setlength{\tabcolsep}{1.5pt}
          \renewcommand{\arraystretch}{1.2}
          \begin{tabular}{lcccc}
            \toprule
            \multirow{2}{*}{Configuration} & \multicolumn{2}{c}{Binary (AUROC $\uparrow$)} & \multicolumn{2}{c}{Multi-class (Kappa $\uparrow$)} \\
            \cmidrule(lr){2-3} \cmidrule(lr){4-5}
            & Mental & CHB-MIT & TUEV & BCICIV-2a \\
            \midrule
            Full Model & \textbf{0.862} & \textbf{0.928} & \textbf{0.690} & \textbf{0.388} \\
            \midrule
            \multicolumn{5}{l}{\textit{(a) Dual-Domain Token.}} \\
            \hspace{1mm} w/o Freq codebook & 0.845 & 0.887 & 0.639 & 0.373 \\
            \hspace{1mm} w/o Time codebook & 0.857 & 0.896 & 0.655 & 0.374 \\
            \midrule
            \multicolumn{5}{l}{\textit{(b) Residual Quant.}} \\
            \hspace{1mm} Single-layer VQ   & 0.849 & 0.896 & 0.682 & 0.321 \\
            \midrule
            \multicolumn{5}{l}{\textit{(c) Hierarchical AutoReg.}} \\
            \hspace{1mm} Independent pred. & 0.780 & 0.875 & 0.662 & 0.335 \\
            \midrule
            \multicolumn{5}{l}{\textit{(d) Masking}} \\
            \hspace{1mm} Random masking   & 0.852 & 0.910 & 0.673 & 0.363 \\
            \bottomrule
          \end{tabular}
        }
    \end{small}
  \end{center}
  \vskip -0.1in
\end{table}

\begin{figure*}[t]
  \vskip 0.2in
  \begin{center}
    \centerline{\includegraphics[width=\textwidth]{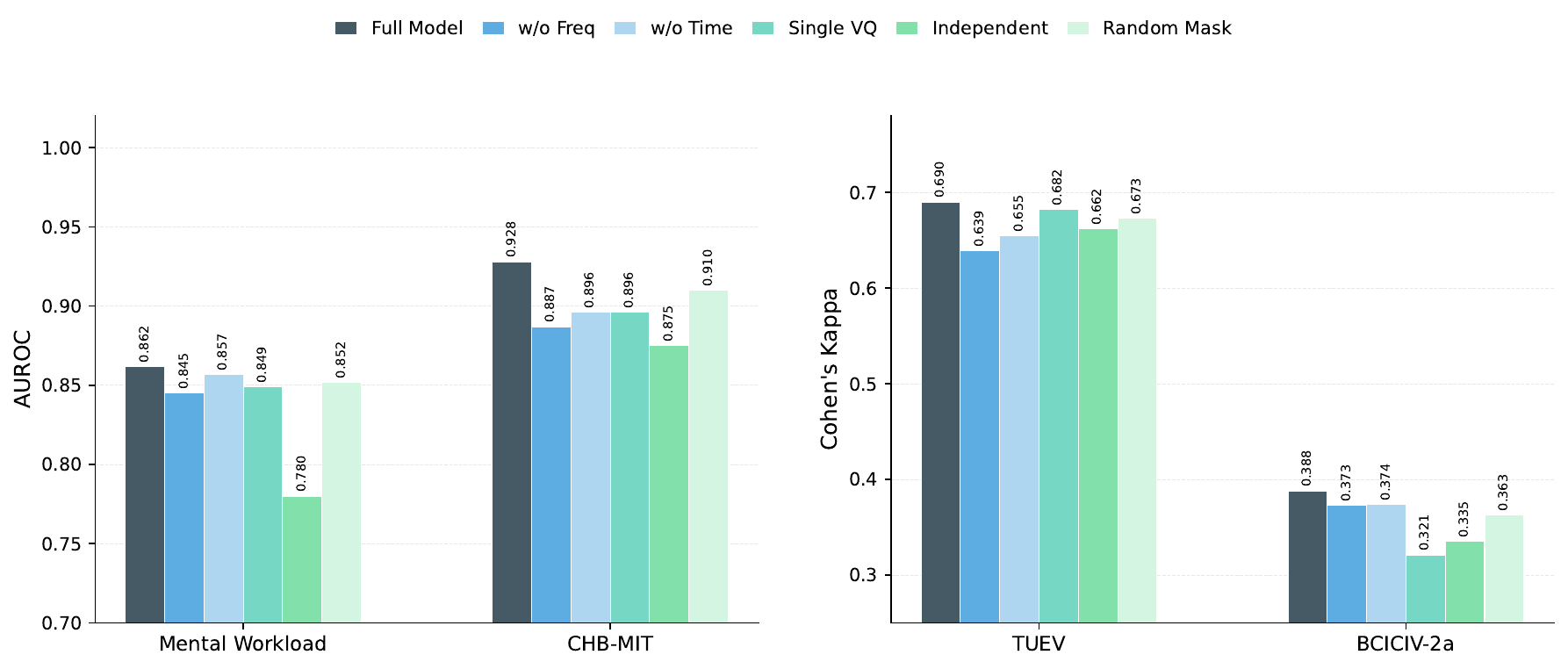}}
    \caption{
      Visualization of ablation study results. The proposed Full Model (dark blue) consistently outperforms single-domain and non-hierarchical variants across all datasets. 
      \textbf{Left:} AUROC scores for Mental Workload and CHB-MIT. 
      \textbf{Right:} Cohen's Kappa scores for TUEV and BCICIV-2a.
    }
    \label{fig:ablation}
  \end{center}
  \vskip -0.2in
\end{figure*}


\textbf{Effectiveness of Dual-Domain Tokenization.} 
The removal of either the temporal or frequency codebook results in consistent performance deterioration, underscoring the complementary nature of these two representations. Specifically, the frequency-only variant (w/o Time) retains competitive performance on Mental Workload ($\Delta$0.6\%) and CHB-MIT ($\Delta$3.4\%), suggesting that cognitive states and seizure anomalies are primarily characterized by spectral power shifts. Conversely, the temporal-only variant (w/o Freq) suffers a more significant drop on TUEV ($\Delta$7.4\%), where clinical event classification relies heavily on waveform morphology. The superiority of the full DD-RVQ across all tasks validates our design of jointly modeling time-frequency dynamics.

\textbf{Impact of Residual Quantization.} 
Collapsing our 3-layer RVQ into a standard single-layer VQ yields an average performance decline of 5.9\%. This degradation is particularly pronounced on BCICIV-2a ($\Delta$17.3\%), a task requiring the discrimination of fine-grained motor imagery signals. We hypothesize that the hierarchical residual codes naturally capture the multi-scale structure of neural dynamics: coarse layers encode general intent, while fine layers resolve subtle class differences. These results confirm that residual quantization provides the necessary representational granularity for complex EEG decoding.

\textbf{Necessity of Hierarchical Autoregression.} 
We replaced the hierarchical autoregressive predictor with an independent prediction mechanism—where all codebook layers are predicted simultaneously without conditioning on coarser codes. This modification causes a substantial 8.3\% average degradation, with severe impacts on Mental Workload ($\Delta$9.5\%) and BCICIV-2a ($\Delta$13.7\%). This finding confirms that explicitly modeling coarse-to-fine dependencies guides the encoder to learn more structured and semantically rich representations, which is critical for downstream tasks involving subtle signal variations.

\textbf{Benefit of information Masking.} 
Ablating the importance-guided curriculum masking in favor of uniform random masking leads to a 3.0\% drop in average performance. The impact is most significant on the data-scarce BCICIV-2a dataset ($\Delta$6.4\%). This validates that prioritizing information-rich neural events during pre-training significantly improves data efficiency and transferability, enabling the model to learn robust features even from limited data.

\section{Conclusion}

In this work, we presented BrainRVQ, a high-fidelity EEG foundation model that unifies temporal and spectral modeling through DD-RVQ. By integrating hierarchical autoregressive pre-training with importance-guided curriculum masking, our framework effectively captures the coarse-to-fine structure of complex neural dynamics. Extensive experiments across eight diverse benchmarks demonstrate that BrainRVQ consistently outperforms state-of-the-art baselines, exhibiting robust generalization in seizure detection, motor imagery, and BCI tasks. We believe BrainRVQ provides a strong foundation for developing practical BCI systems and clinical EEG analysis tools. Future work includes scaling to larger models and incorporating multi-modal physiological signals.

\section*{Impact Statement} This work contributes to the advancement of machine learning for physiological signal processing. By developing robust foundation models for EEG, our research has the potential to accelerate the development of more accurate clinical diagnostic tools and efficient assistive neurotechnologies, ultimately benefiting healthcare and human-machine interaction domains.

\nocite{langley00}

\bibliography{reference}
\bibliographystyle{icml2026}

\newpage
\appendix
\onecolumn

\section{Related Work}
\label{app:related_work}

\subsection{EEG Decoding Methods}

EEG is a non-invasive technique to measure brain activity. Early studies on EEG decoding predominantly employed traditional machine learning methods~\cite{bashashati2007survey, lotte2007review}, which usually depend on hand-crafted features that require extensive prior knowledge and often exhibit weak generalizability. With the development of deep learning techniques, an increasing number of researchers have shifted their focus to studying EEG decoding methods based on deep learning~\cite{craik2019deep, roy2019deep}.

Convolutional neural networks (CNNs) have been widely adopted to extract temporal and spatial features from EEG for various BCI tasks, including motor imagery classification, emotion recognition, and seizure detection~\cite{schirrmeister2017deep, lawhern2018eegnet}. Long Short-Term Memory (LSTM) networks have also been employed for EEG feature extraction and classification on tasks such as motor imagery and sleep staging~\cite{supratak2017deepsleepnet}. Transformer architectures have been utilized to learn spatial-temporal features for BCI tasks including emotion recognition, sleep staging, and person identification~\cite{song2021transformer, lee2022eeg}. To combine the strengths of CNN and Transformer, some works devise CNN-Transformer hybrid networks for EEG classification~\cite{ma2022novel}.

\subsection{EEG Foundation Models}

Foundation models~\cite{wiggins2022opportunities}, such as BERT~\cite{devlin2019bert}, MAE~\cite{he2022masked}, and GPT-4~\cite{achiam2023gpt}, have achieved remarkable success in computer vision and natural language processing. However, the potential of foundation models for brain signals remains largely unexplored.

Recent efforts have begun to address this gap. BENDR~\cite{kostas2021bendr} employs contrastive self-supervised learning to learn generic EEG representations, addressing the problem of limited labeled data. BrainBERT~\cite{wang2023brainbert} is a pre-training model for intracranial recordings that learns complex non-linear transformations through masked spectrogram modeling. Brant~\cite{zhang2023brant} proposes a foundation model for intracranial neural signals that captures long-term dependencies and spatial correlations across channels.

BIOT~\cite{yang2023biot} introduces a generic biosignal learning model that enables joint pre-training and knowledge transfer across different biosignal datasets. LaBraM~\cite{jiang2024large} presents a large brain model that learns generic EEG representations by predicting neural tokens of masked EEG patches through a VQ-VAE-based tokenizer. EEGPT~\cite{wang2024eegpt} proposes a mask-based dual self-supervised learning method for efficient feature extraction. CBraMod~\cite{wang2024cbramod} introduces criss-cross attention to separately model spatial and temporal dependencies. CodeBrain~\cite{ma2025codebrain} explores state-space models with dual-domain tokenization for EEG representation learning.

Despite these advances, existing approaches often rely on single-domain tokenization and flat vector quantization, limiting their ability to capture the hierarchical structure of neural dynamics. Our work addresses these limitations through dual-domain residual vector quantization and hierarchical autoregressive pre-training.

\subsection{Vector Quantization in Neural Signal Processing}

Vector Quantization (VQ) has emerged as a powerful technique for learning discrete representations of continuous signals. VQ-VAE~\cite{van2017neural} introduced the concept of learning discrete latent codes through vector quantization, enabling effective compression and generation of complex data.

RVQ~\cite{zeghidour2021soundstream} extends this approach by employing a cascade of codebooks to progressively refine the representation. This hierarchical quantization has proven particularly effective in high-fidelity audio generation, as demonstrated by SoundStream~\cite{zeghidour2021soundstream}, EnCodec~\cite{defossez2022high}, and AudioLM~\cite{borsos2023audiolm}.

In the context of EEG, LaBraM~\cite{jiang2024large} first introduced neural tokenizers for discretizing EEG signals, but employed single-layer VQ in the time domain only. Our DD-RVQ approach extends this by performing hierarchical quantization in both time and frequency domains, enabling more comprehensive capture of spectro-temporal neural dynamics.

\section{Preliminaries}
\label{app:method_details}

\subsection{Codebook Learning with Exponential Moving Average}

To ensure stable codebook learning and prevent mode collapse, we employ Exponential Moving Average (EMA) updates for the codebook vectors instead of gradient-based optimization. For each codebook $\mathbf{V}^{(l)} = \{\mathbf{v}_k^{(l)}\}_{k=1}^{K}$ at level $l$, the update rule at training step $t$ is:

\begin{equation}
\mathbf{n}_k^{(t)} = \gamma \cdot \mathbf{n}_k^{(t-1)} + (1 - \gamma) \cdot \sum_{i} \mathbf{1}[z_i^{(l)} = k]
\end{equation}
\begin{equation}
\mathbf{m}_k^{(t)} = \gamma \cdot \mathbf{m}_k^{(t-1)} + (1 - \gamma) \cdot \sum_{i: z_i^{(l)} = k} \mathbf{e}_i
\end{equation}
\begin{equation}
\mathbf{v}_k^{(l)} \leftarrow \frac{\mathbf{m}_k^{(t)}}{\mathbf{n}_k^{(t)} + \epsilon}
\end{equation}

where $\gamma = 0.99$ is the decay rate, $\mathbf{n}_k$ tracks the exponential moving count of assignments to code $k$, $\mathbf{m}_k$ tracks the exponential moving sum of assigned embeddings, and $\epsilon = 10^{-6}$ prevents division by zero. This EMA approach provides several advantages:
\begin{itemize}
    \item \textbf{Stability}: Gradual updates prevent abrupt codebook changes that can destabilize training.
    \item \textbf{Utilization}: Combined with $\ell_2$-normalization of embeddings before quantization, EMA encourages uniform codebook utilization.
    \item \textbf{Efficiency}: No gradient computation is required for codebook updates.
\end{itemize}

Additionally, we apply $\ell_2$-normalization to both encoder outputs and codebook vectors before computing distances:
\begin{equation}
\tilde{\mathbf{e}} \leftarrow \frac{\tilde{\mathbf{e}}}{\|\tilde{\mathbf{e}}\|_2}, \quad \mathbf{v}_k \leftarrow \frac{\mathbf{v}_k}{\|\mathbf{v}_k\|_2}
\end{equation}
This normalization projects all vectors onto a unit hypersphere, making the codebook learning purely directional and improving training stability.

\subsection{Discrete Fourier Transform for EEG Analysis}
Given an EEG patch $\mathbf{x} \in \mathbb{R}^P$, the Discrete Fourier Transform (DFT) computes its spectral representation:
\begin{equation}
    X[k] = \sum_{n=0}^{P-1} x[n] \cdot e^{-j\frac{2\pi kn}{P}}, \quad k = 0, 1, \ldots, P-1
\end{equation}
From the complex-valued spectrum, we extract amplitude and phase:
\begin{align}
    A[k] &= |X[k]| = \sqrt{\text{Re}(X[k])^2 + \text{Im}(X[k])^2} \\
    \phi[k] &= \arg(X[k]) = \arctan\left(\frac{\text{Im}(X[k])}{\text{Re}(X[k])}\right)
\end{align}
The amplitude spectrum captures the power distribution across frequency bands (e.g., delta: 0.5-4 Hz, theta: 4-8 Hz, alpha: 8-13 Hz, beta: 13-30 Hz), while the phase spectrum encodes temporal relationships critical for neural synchronization patterns.

\subsection{Physiology-Aware Importance Scoring}

This section provides the detailed formulation of the importance scoring function used in our curriculum masking strategy.

\subsubsection{Frequency-Domain Metrics}

Given a patch $\bar{x}_{c,a} \in \mathbb{R}^{P}$, we first compute its power spectral density via the Discrete Fourier Transform:
\begin{equation}
    \mathbf{P}[k] = |X[k]|^2, \quad k = 0, 1, \ldots, \lfloor P/2 \rfloor
\end{equation}
where $X[k]$ is the DFT coefficient at frequency bin $k$. The corresponding frequency for each bin is $f_k = k \cdot f_s / P$, where $f_s$ is the sampling rate.

The \textbf{neural band ratio} measures the proportion of power within physiologically relevant frequency bands:
\begin{equation}
    S_{\text{neural}} = \frac{\sum_{k: f_k \in [4, 30)\text{Hz}} \mathbf{P}[k]}{\sum_k \mathbf{P}[k] + \epsilon}
\end{equation}
This range encompasses theta (4--8\,Hz), alpha (8--13\,Hz), and beta (13--30\,Hz) rhythms, which are strongly associated with cognitive processes and clinical biomarkers.

The \textbf{artifact penalty} quantifies contamination from non-neural sources:
\begin{equation}
    S_{\text{clean}} = 1 - \frac{\sum_{k: f_k < 2\text{Hz} \lor f_k \geq 45\text{Hz}} \mathbf{P}[k]}{\sum_k \mathbf{P}[k] + \epsilon}
\end{equation}
Low-frequency components ($<$2\,Hz) typically reflect baseline drift and electrode artifacts, while high-frequency components ($\geq$45\,Hz) often correspond to electromyographic (muscle) contamination.

\subsubsection{Time-Domain Metrics via Hjorth Parameters}

Hjorth parameters~\citep{hjorth1970eeg} provide computationally efficient estimates of signal dynamics. Let $\Delta \bar{x}[p] = \bar{x}[p] - \bar{x}[p-1]$ and $\Delta^2 \bar{x}[p] = \Delta \bar{x}[p] - \Delta \bar{x}[p-1]$ denote the first and second-order differences.

The \textbf{activity} parameter captures signal variance:
\begin{equation}
    S_{\text{activity}} = \log(\text{Var}(\bar{x}_{c,a}) + \epsilon)
\end{equation}
The logarithmic transformation reduces sensitivity to high-amplitude artifacts.

The \textbf{mobility} parameter approximates the mean frequency:
\begin{equation}
    S_{\text{mobility}} = \sqrt{\frac{\text{Var}(\Delta \bar{x}_{c,a})}{\text{Var}(\bar{x}_{c,a}) + \epsilon}}
\end{equation}

The \textbf{complexity} parameter measures spectral bandwidth:
\begin{equation}
    S_{\text{complexity}} = \frac{1}{S_{\text{mobility}} + \epsilon} \sqrt{\frac{\text{Var}(\Delta^2 \bar{x}_{c,a})}{\text{Var}(\Delta \bar{x}_{c,a}) + \epsilon}}
\end{equation}
High complexity indicates rich multi-frequency content, whereas low values correspond to narrow-band or monotonic signals.

\subsubsection{Irregularity Metric}

To capture non-stationarity, we compute an irregularity metric based on the variability of signal changes:
\begin{equation}
    S_{\text{irreg}} = \frac{\mathbb{E}[|\Delta|\Delta \bar{x}_{c,a}||]}{\mathbb{E}[|\Delta \bar{x}_{c,a}|] + \epsilon}
\end{equation}
This measures how consistently the signal changes over time, with higher values indicating more irregular (and potentially more informative) patterns.

\subsubsection{Score Aggregation}

The final importance score aggregates all metrics via weighted combination:
\begin{equation}
    S_{c,a} = \alpha_1 \tilde{S}_{\text{neural}} + \alpha_2 \tilde{S}_{\text{clean}} + \alpha_3 \tilde{S}_{\text{complexity}} + \alpha_4 \tilde{S}_{\text{irreg}} + \alpha_5 \tilde{S}_{\text{mobility}}
\end{equation}
where $\tilde{S}_m$ denotes min-max normalization across all patches within a sample. In our implementation, we use $\alpha_1 = 0.30$, $\alpha_2 = 0.25$, $\alpha_3 = 0.20$, $\alpha_4 = 0.15$, and $\alpha_5 = 0.10$, prioritizing neural band content and artifact-free segments.

\subsubsection{Curriculum Weight Schedule}

The curriculum weight $w(t)$ controls the balance between importance-guided and random masking:
\begin{equation}
    w(t) = w_0 + (w_{\max} - w_0) \cdot \frac{t}{T_{\text{total}}}
\end{equation}
We set $w_0 = 0.2$ and $w_{\max} = 0.7$, ensuring that early training explores diverse patches while later training focuses on information-rich regions. The temperature parameter for softmax sampling is set to $\tau = 0.8$.

\section{Pre-training Analysis}
\label{app:pretrain_analysis}

This section provides a detailed analysis of the pre-training process for BrainRVQ. We first present the complete hyperparameter configurations for reproducibility. Then, we analyze the training dynamics, including loss convergence curves and codebook utilization patterns for both the DD-RVQ tokenizer and hierarchical autoregressive pre-training stages. Interpretability analysis and qualitative visualizations of the learned representations are provided separately.

\subsection{Pre-training Settings}
Here we provide complete hyperparameter configurations for both the DD-RVQ tokenizer and hierarchical autoregressive pre-training stages in Table~\ref{tab:pretrain_hyperparams}. 

\begin{table}[htbp!]
\centering
\caption{Hyperparameters for BrainRVQ pre-training.}
\label{tab:pretrain_hyperparams}
\begin{tabular}{lcc}
\toprule
\textbf{Hyperparameters} & \textbf{Tokenizer} & \textbf{Pre-training} \\
\midrule
\multicolumn{3}{l}{\textit{EEG Sample}} \\
\quad Channels & \multicolumn{2}{c}{19} \\
\quad Time points & \multicolumn{2}{c}{6000 (30 seconds @ 200 Hz)} \\
\quad Patch dimension & \multicolumn{2}{c}{200 (1 second)} \\
\quad Sequence length & \multicolumn{2}{c}{$19 \times 30 = 570$} \\
\quad Mask ratio & -- & 0.5 \\
\quad Mask token & -- & Learnable \\
\midrule
\multicolumn{3}{l}{\textit{Patch Encoder (3-layer 2D Temporal Conv)}} \\
\quad Input channels & \multicolumn{2}{c}{\{1, 8, 8\}} \\
\quad Output channels & \multicolumn{2}{c}{\{8, 8, 8\}} \\
\quad Kernel size & \multicolumn{2}{c}{\{(1,15), (1,3), (1,3)\}} \\
\quad Stride & \multicolumn{2}{c}{\{(1,8), (1,1), (1,1)\}} \\
\quad Padding & \multicolumn{2}{c}{\{(0,7), (0,1), (0,1)\}} \\
\midrule
\multicolumn{3}{l}{\textit{Residual Vector Quantization (per domain)}} \\
\quad RVQ layers & 3 & -- \\
\quad Codebook sizes & [8192, 8192, 8192] & -- \\
\quad Code embedding dimension & 64 & -- \\
\quad EMA decay & 0.99 & -- \\
\quad Commitment weight ($\beta$) & 1.0 & -- \\
\midrule
\multicolumn{3}{l}{\textit{Transformer Encoder}} \\
\quad Layers & 12 & 12 \\
\quad Hidden dimension & 200 & 200 \\
\quad Attention heads & 10 & 10 \\
\quad Feed-forward dimension & 800 & 800 \\
\quad Drop path rate & 0.0 & 0.1 \\
\midrule
\multicolumn{3}{l}{\textit{Transformer Decoder}} \\
\quad Layers & 3 & -- \\
\quad Hidden dimension & 200 & -- \\
\quad Attention heads & 10 & -- \\
\midrule
\multicolumn{3}{l}{\textit{Importance-Guided Masking}} \\
\quad Softmax temperature ($\tau$) & -- & 0.8 \\
\quad Final importance weight & -- & 0.7 \\
\quad Symmetric loss & -- & True \\
\midrule
\multicolumn{3}{l}{\textit{Training}} \\
\quad Epochs & 20 & 20 \\
\quad Batch size & 128 & 64 \\
\quad Optimizer & AdamW & AdamW \\
\quad Learning rate & $5 \times 10^{-4}$ & $5 \times 10^{-4}$ \\
\quad Adam $\beta$ & (0.9, 0.999) & (0.9, 0.999) \\
\quad Adam $\epsilon$ & $1 \times 10^{-8}$ & $1 \times 10^{-8}$ \\
\quad Weight decay & $5 \times 10^{-2}$ & $5 \times 10^{-2}$ \\
\quad Scheduler & CosineAnnealingLR & CosineAnnealingLR \\
\quad Warmup epochs & 5 & 5 \\
\quad Minimal learning rate & $1 \times 10^{-5}$ & $1 \times 10^{-5}$ \\
\bottomrule
\end{tabular}
\end{table}

\subsection{Training Dynamics}

We analyze the convergence behavior of both training stages to provide insights into the learning process of BrainRVQ.

\begin{figure}[htbp]
  \centering
  \includegraphics[width=1\columnwidth]{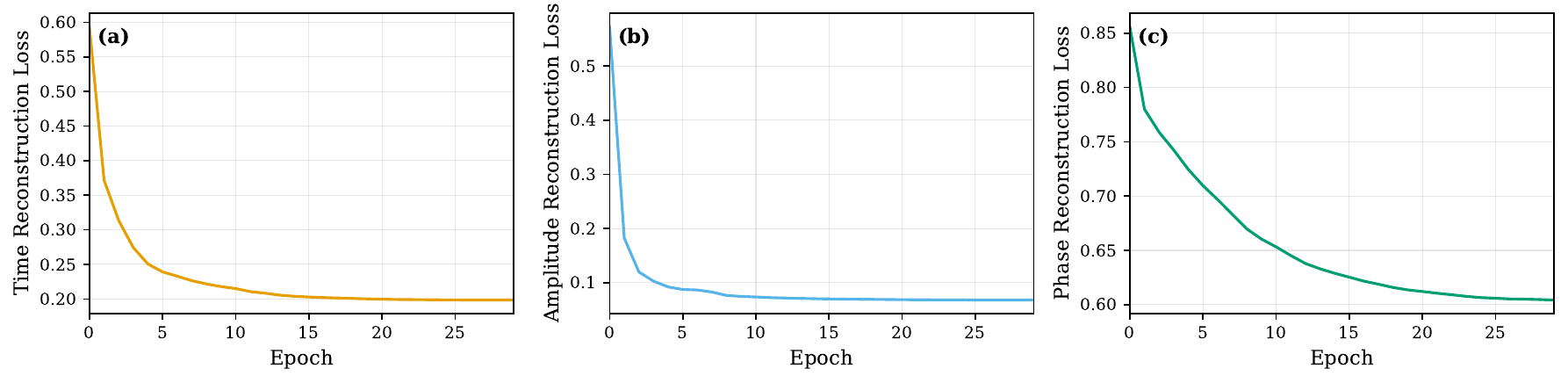}
  \caption{DD-RVQ tokenizer training dynamics. The total loss (black) comprises time-domain waveform reconstruction (blue), frequency-domain amplitude reconstruction (orange), and phase reconstruction (green).}
  \label{fig:ddrvq_loss}
\end{figure}

\textbf{DD-RVQ Tokenizer Training.}
Figure~\ref{fig:ddrvq_loss} shows the loss curves during DD-RVQ training. The total loss comprises three reconstruction objectives: time-domain waveform, frequency-domain amplitude, and phase reconstruction. We observe rapid convergence during the first 20 epochs, followed by gradual stabilization. Among the three components, amplitude reconstruction converges fastest due to the relatively smooth and predictable nature of spectral power distributions. Time-domain reconstruction exhibits slower convergence, reflecting the higher complexity of raw waveform modeling. Phase reconstruction shows higher variance throughout training, consistent with findings in audio processing where phase is inherently more challenging to reconstruct than magnitude~\citep{zeghidour2021soundstream}.

\begin{figure}[htbp]
  \centering
  \includegraphics[width=0.6\columnwidth]{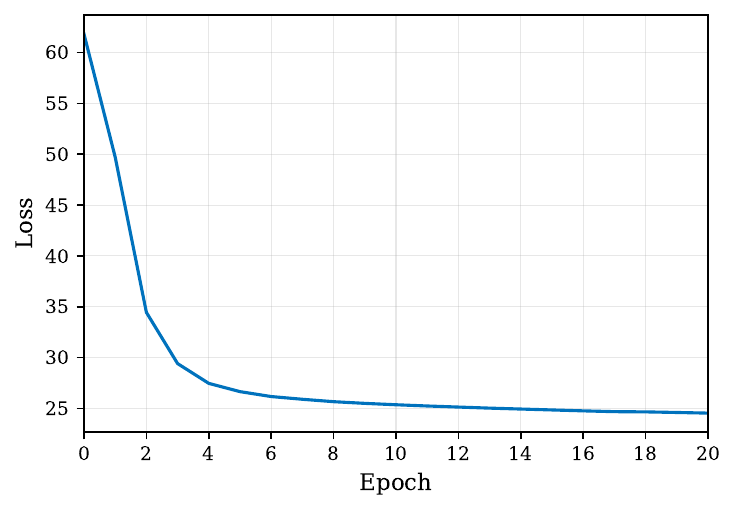}
  \caption{Pre-training loss dynamics. Layer-0 (coarse) achieves consistently lower loss than Layer-1/2 (fine), validating the coarse-to-fine learning hierarchy.}
  \label{fig:pretrain_loss}
\end{figure}

\begin{figure}[htbp]
  \centering
  \includegraphics[width=0.6\columnwidth]{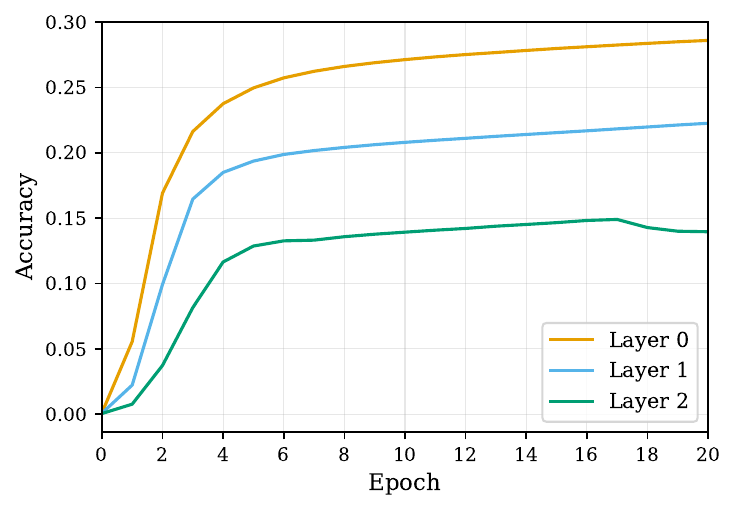}
  \caption{Layer-wise token prediction accuracy during pre-training. Layer-0 achieves the highest accuracy ($\sim$29\%), followed by Layer-1 ($\sim$22\%) and Layer-2 ($\sim$14\%), confirming that coarse codes are more predictable than fine-grained residuals.}
  \label{fig:pretrain_acc}
\end{figure}

\textbf{Hierarchical Autoregressive Pre-training.}
We examine both loss curves (Figure~\ref{fig:pretrain_loss}) and prediction accuracy (Figure~\ref{fig:pretrain_acc}) during pre-training to validate our hierarchical design. As shown in Figure~\ref{fig:pretrain_loss}, we observe distinct convergence patterns across RVQ layers: Layer-0 prediction loss is consistently lower than Layer-1 and Layer-2, confirming that coarse codes capturing global patterns (e.g., dominant frequency bands, overall amplitude envelope) are easier to predict than fine-grained residual codes. The stable loss gap between adjacent layers after warmup validates our layer-wise weighting scheme ($\lambda_0 > \lambda_1 > \lambda_2$).

Figure~\ref{fig:pretrain_acc} further corroborates this observation from the accuracy perspective. Layer-0 achieves the highest prediction accuracy of approximately 29\%, while Layer-1 and Layer-2 reach 22\% and 14\%, respectively. This monotonic decrease in accuracy across layers aligns with our intuition: coarse-level codes encode dominant signal characteristics that exhibit stronger temporal dependencies and are thus more predictable, whereas fine-level residual codes capture high-frequency details and noise-like components that are inherently harder to anticipate. Notably, all layers show rapid improvement within the first 5 epochs before gradually stabilizing, indicating efficient learning of the hierarchical token structure.

\subsection{Codebook Utilization}

\begin{figure}[htbp]
    \centering
    \includegraphics[width=0.9\textwidth]{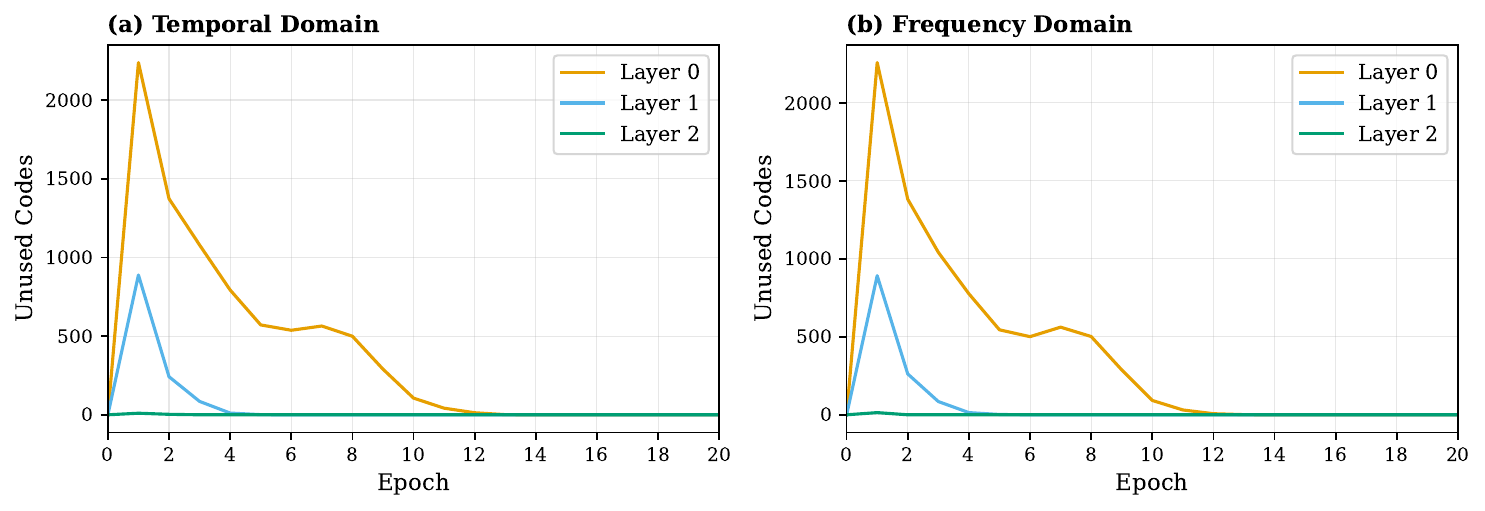}
    \caption{Number of unused codes during DD-RVQ training for (a) temporal and (b) frequency codebooks. All layers converge to near-zero unused codes within 20 epochs, achieving 100\% codebook utilization.}
    \label{fig:unused_codes}
\end{figure}

\begin{figure}[htbp]
    \centering
    \includegraphics[width=0.95\textwidth]{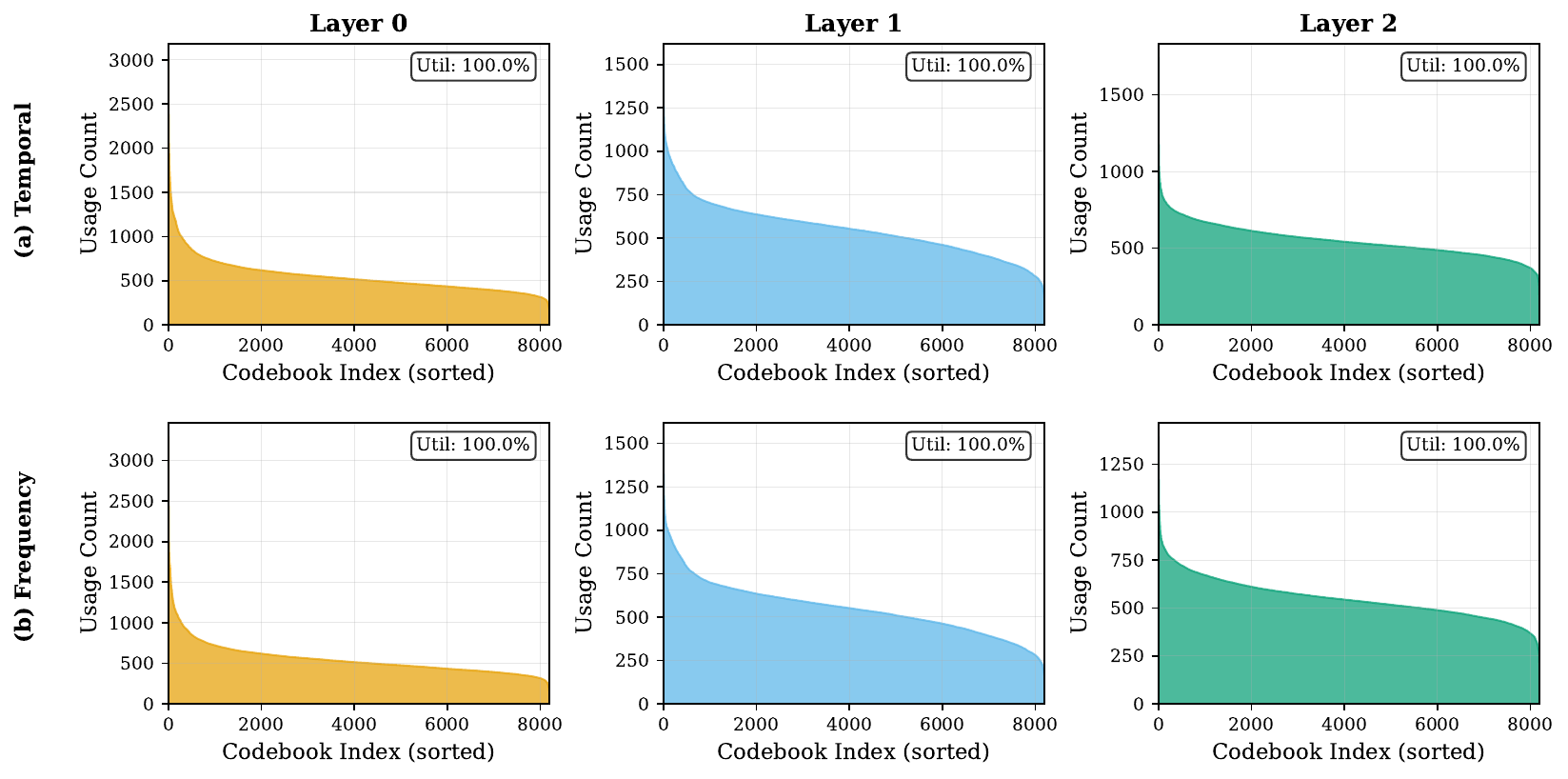}
    \caption{Final codebook usage distribution for (a) temporal and (b) frequency domains across three RVQ layers. All codebooks achieve 100\% utilization with near-uniform distributions (normalized entropy $>$0.99).}
    \label{fig:codebook_dist}
\end{figure}

Codebook utilization is a critical indicator of effective discrete representation learning. Under-utilized codebooks suffer from ``codebook collapse,'' where only a small subset of codes are actively used while the majority remain dormant, severely limiting representational capacity. We analyze our DD-RVQ tokenizer from two perspectives: training dynamics of unused codes (Figure~\ref{fig:unused_codes}) and final code frequency distributions (Figure~\ref{fig:codebook_dist}).

\textbf{Training Dynamics.}
Figure~\ref{fig:unused_codes} tracks the number of unused codes throughout training. Initially, approximately 2,000 codes remain dormant across all layers. As training progresses, this number decreases rapidly, with all six codebooks achieving near-complete utilization within 20 epochs. Notably, the frequency codebook (Figure~\ref{fig:unused_codes}b) demonstrates slightly faster convergence compared to the temporal codebook (Figure~\ref{fig:unused_codes}a), which we attribute to the smoother and more structured nature of spectral representations. Across RVQ layers, Layer-0 (coarse) converges fastest, followed by Layer-1 and Layer-2, consistent with the intuition that dominant patterns are captured earlier during training.

\textbf{Final Distribution Analysis.}
Figure~\ref{fig:codebook_dist} visualizes the code frequency distributions after training, sorted by usage count. All six codebooks achieve 100\% utilization, indicating complete absence of codebook collapse. The distributions exhibit near-uniform characteristics, as quantified in Table~\ref{tab:codebook_stats}.

The normalized entropy exceeds 0.99 for all codebooks (where 1.0 indicates perfect uniformity), demonstrating highly balanced code utilization. Second, the Gini coefficients are consistently low (0.10--0.17), further confirming the absence of dominant codes monopolizing the representation space. Third, deeper RVQ layers exhibit more uniform distributions (Layer-2: Gini=0.104) compared to shallower layers (Layer-0: Gini=0.174), suggesting that residual quantization progressively distributes information more evenly across the codebook. Fourth, the Top-10\% contribution metric shows that the most frequent 10\% of codes account for only 13.7--18.1\% of total usage, close to the ideal value of 10\% under uniform distribution. These results validate that our DD-RVQ tokenizer effectively utilizes the full codebook capacity without collapse, providing a rich discrete vocabulary for representing diverse EEG patterns.

\begin{table}[h]
\centering
\caption{Codebook utilization statistics. Normalized entropy close to 1.0 and low Gini coefficients indicate near-uniform code usage across all layers.}
\label{tab:codebook_stats}
\setlength{\tabcolsep}{8pt}
\begin{tabular}{lccc}
\toprule
\textbf{Codebook} & \textbf{Norm. Entropy} & \textbf{Gini Coef.} & \textbf{Top-10\% Contrib.} \\
\midrule
Temporal Layer-0 & 0.994 & 0.174 & 18.1\% \\
Temporal Layer-1 & 0.996 & 0.151 & 15.4\% \\
Temporal Layer-2 & 0.998 & 0.104 & 13.8\% \\
\midrule
Frequency Layer-0 & 0.994 & 0.174 & 18.1\% \\
Frequency Layer-1 & 0.996 & 0.151 & 15.5\% \\
Frequency Layer-2 & 0.998 & 0.104 & 13.7\% \\
\bottomrule
\end{tabular}
\end{table}

\section{More Results on Downstream BCI Tasks}
\label{app:downstream}

In this section, we provide more details for experimental settings, including dataset descriptions, baseline methods, evaluation metrics, and extended results on additional datasets not presented in the main text.

\subsection{Fine-tuning Settings}

For downstream evaluation, we load the pre-trained encoder weights and attach a task-specific classification head consisting of global average pooling followed by a linear projection layer. We fine-tune the entire model end-to-end using AdamW optimizer with layer-wise learning rate decay (decay factor 0.65) to prevent catastrophic forgetting of pre-trained representations, where earlier layers receive progressively smaller learning rates than later layers. We use cross-entropy loss for all classification tasks. Table~\ref{tab:finetune_hyperparams} summarizes the hyperparameters shared across all downstream tasks.

For downstream evaluation, we load the pre-trained encoder weights and attach a task-specific classification head consisting of a global average pooling layer followed by a three-layer Multi-Layer Perceptron (MLP). We fine-tune the entire model end-to-end using AdamW optimizer with layer-wise learning rate decay (decay factor 0.65) to prevent catastrophic forgetting of pre-trained representations, where earlier layers receive progressively smaller learning rates than later layers. We use cross-entropy loss for all classification tasks. Table~\ref{tab:finetune_hyperparams} summarizes the hyperparameters shared across all downstream tasks.

\begin{table}[h]
\centering
\caption{Fine-tuning hyperparameters for downstream tasks.}
\label{tab:finetune_hyperparams}
\begin{tabular}{lc}
\toprule
\textbf{Hyperparameter} & \textbf{Value} \\
\midrule
Optimizer & AdamW \\
Adam $\beta$ & (0.9, 0.999) \\
Adam $\epsilon$ & $1 \times 10^{-8}$ \\
Learning rate & $5 \times 10^{-4}$ \\
Weight decay & $5 \times 10^{-2}$ \\
Layer decay & 0.65 \\
Drop path rate & 0.1 \\
Batch size & 64 \\
Warmup epochs & 5 \\
Total epochs & 50 \\
Scheduler & Cosine annealing \\
\bottomrule
\end{tabular}
\end{table}

\subsection{Downstream Datasets}

We evaluate BrainRVQ on 8 downstream BCI tasks using publicly available datasets. For all downstream datasets, we resample the EEG signals to 200 Hz and set the patch duration to 1 second, consistent with the pre-training configuration. 

\textbf{TUAB}~\citep{obeid2016temple} is a clinical EEG corpus from the Temple University Hospital, annotated as normal or abnormal by certified neurologists. The EEG signals are originally recorded at 23 channels with a sampling rate of 256 Hz. Following prior work~\citep{yang2023biot,wang2024cbramod}, we select 16 channels based on the bipolar montage in the international 10-20 system. A bandpass filter (0.3--75 Hz) is applied to remove low-frequency drift and high-frequency noise, and a notch filter at 60 Hz is used to suppress power line interference. All signals are resampled to 200 Hz and segmented into 10-second non-overlapping windows, yielding 409,455 samples for binary classification. We follow the official train/test split and further divide the training subjects into training and validation sets with an 8:2 ratio.

\textbf{TUEV}~\citep{obeid2016temple} is a clinical EEG corpus containing annotations across six event types: spike and sharp wave (SPSW), generalized periodic epileptiform discharges (GPED), periodic lateralized epileptiform discharges (PLED), eye movement (EYEM), artifact (ARTF), and background (BCKG). We adopt the same preprocessing protocol as TUAB, selecting 16 bipolar montage channels with identical filtering. All signals are resampled to 200 Hz and segmented into 5-second windows, yielding 112,491 samples. We follow the official train/test split and further divide the training set into 80\%:20\% for training and validation.

\textbf{CHB-MIT}~\citep{shoeb2009application} is a pediatric seizure detection dataset collected at Children's Hospital Boston, consisting of long-term EEG recordings from 23 subjects with intractable epilepsy. The signals were originally recorded at 256 Hz. Following prior work~\citep{yang2023biot,wang2024cbramod}, we select 16 channels based on the international 10-20 system. All signals are resampled to 200 Hz and segmented into 10-second non-overlapping windows, yielding 326,993 samples. Subjects 1--19 are used for training, subjects 20--21 for validation, and subjects 22--23 for testing. This dataset is highly imbalanced, with seizure events constituting less than 5\% of the total samples.

\textbf{SEED-V}~\citep{liu2021comparing} is an emotion recognition dataset containing EEG recordings from 16 subjects watching video clips designed to elicit five emotional states: happy, sad, neutral, disgust, and fear. EEG signals are recorded at 62 channels with a sampling rate of 1000 Hz. The signals are downsampled to 200 Hz and segmented into 1-second windows, yielding 117,744 samples. Each subject participates in 3 sessions with 15 trials per session. Following prior work~\citep{wang2024cbramod}, we divide the 15 trials evenly into training (5), validation (5), and test (5) sets for session-wise evaluation.

\textbf{PhysioNet-MI}~\citep{schalk2004bci2000} is a motor imagery dataset from the BCI2000 system, collected from 109 subjects using 64 channels at 160 Hz. The dataset contains four motor imagery classes: left fist, right fist, both fists, and both feet. Following standard practice, the EEG signals are segmented using the [2, 6] second window relative to cue onset and resampled to 200 Hz, yielding 9,837 4-second samples. Subjects 1--70 are used for training, subjects 71--89 for validation, and subjects 90--109 for testing.

\textbf{SHU-MI}~\citep{malarge} is a large-scale motor imagery dataset designed for studying cross-session variability. It contains 32-channel EEG recordings at 250 Hz from 25 subjects performing binary motor imagery tasks (left hand vs. right hand). The signals are resampled to 200 Hz and segmented into 4-second windows, yielding 11,988 samples. Subjects 1--15 are used for training, subjects 16--20 for validation, and subjects 21--25 for testing.

\textbf{BCICIV-2a}~\citep{tangermann2012review} is a 4-class motor imagery dataset from BCI Competition IV, containing 22-channel EEG from 9 subjects at 250 Hz. The four classes correspond to imagination of movement of the left hand, right hand, both feet, and tongue. The signals are bandpass filtered (4--40 Hz), resampled to 200 Hz, and segmented using the [2, 6] second window relative to cue onset, yielding 5,088 4-second samples. Subjects 1--5 are used for training, subjects 6--7 for validation, and subjects 8--9 for testing.

\textbf{Mental Workload}~\citep{zyma2019electroencephalograms} is a dataset for cognitive load detection during mental arithmetic tasks. EEG signals are recorded from 36 subjects using 20 channels at 500 Hz. Recordings labeled as ``no stress'' correspond to baseline rest periods, while ``stress'' labels are assigned to recordings during active mental arithmetic. The signals are bandpass filtered (0.5--45 Hz), resampled to 200 Hz, and segmented into 5-second windows, yielding 1,707 samples. Subjects 1--28 are used for training, subjects 29--32 for validation, and subjects 33--36 for testing.

\subsection{Baselines}

We compare BrainRVQ with both supervised baselines and EEG foundation models for comprehensive evaluation. Supervised baselines include \textbf{EEGNet}~\cite{lawhern2018eegnet}, a compact CNN based on depthwise and separable convolutions; \textbf{ST-Transformer}~\cite{song2021transformer}, which utilizes self-attention mechanisms to capture spatial correlations across channels and temporal dependencies over time; For foundation model baselines, we include \textbf{BENDR}~\cite{kostas2021bendr}, which uses contrastive self-supervised learning to learn generic EEG representations; \textbf{BIOT}~\cite{yang2023biot}, a biosignal transformer that enables cross-dataset knowledge transfer; \textbf{LaBraM}~\cite{jiang2024large}, which learns representations by predicting neural tokens of masked EEG patches; and \textbf{CBraMod}~\cite{wang2024cbramod}, which employs criss-cross attention to separately model spatial and temporal dependencies. We fine-tune all foundation models using their publicly released pre-trained weights and follow their official fine-tuning protocols. 

\subsection{Metrics}

We adopt evaluation metrics consistent with prior work~\cite{jiang2024large, wang2024cbramod}. For binary classification tasks (TUAB, CHB-MIT, SHU-MI, Mental Arithmetic), we use \textbf{Balanced Accuracy}, \textbf{AUC-PR} (Area Under Precision-Recall Curve), and \textbf{AUROC} (Area Under ROC Curve), where AUROC serves as the primary metric for model selection. For multi-class classification tasks (TUEV, SEED-V, PhysioNet-MI, BCICIV-2a), we use \textbf{Balanced Accuracy}, \textbf{Cohen's Kappa}, and \textbf{Weighted F1}, where Cohen's Kappa serves as the primary metric. All results are obtained with five different random seeds and reported as mean $\pm$ standard deviation.

\subsection{Results on Additional Downstream Tasks}

In this section, we report detailed results on four downstream tasks not presented in the main text: motor imagery (SHU-MI, PhysioNet-MI), emotion recognition (SEED-V), and abnormal detection (TUAB).

\subsubsection{Motor Imagery Classification (SHU-MI)}

\begin{table}[h]
\centering
\caption{Results on motor imagery classification (SHU-MI, 2-class). Best results are in \textbf{bold}, second best are \underline{underlined}.}
\label{tab:shumi_results}
\begin{tabular}{lccc}
\toprule
\textbf{Method} & \textbf{Bal. Acc} & \textbf{AUC-PR} & \textbf{AUROC} \\
\midrule
EEGNet & 0.589 $\pm$ 0.018 & 0.631 $\pm$ 0.014 & 0.628 $\pm$ 0.015 \\
ST-Transformer & 0.599 $\pm$ 0.021 & 0.639 $\pm$ 0.012 & 0.643 $\pm$ 0.011 \\
BENDR & 0.557 $\pm$ 0.023 & 0.585 $\pm$ 0.027 & 0.586 $\pm$ 0.028 \\
BIOT & 0.618 $\pm$ 0.018 & 0.677 $\pm$ 0.012 & 0.661 $\pm$ 0.013 \\
LaBraM & 0.617 $\pm$ 0.019 & 0.676 $\pm$ 0.008 & 0.660 $\pm$ 0.009 \\
CBraMod & \textbf{0.637 $\pm$ 0.015} & \underline{0.714 $\pm$ 0.009} & \underline{0.699 $\pm$ 0.007} \\
\textbf{\modelname{} (Ours)} & \textbf{0.637 $\pm$ 0.018} & \textbf{0.723 $\pm$ 0.011} & \textbf{0.717 $\pm$ 0.007} \\
\bottomrule
\end{tabular}
\end{table}

Table~\ref{tab:shumi_results} shows the results on SHU-MI binary motor imagery classification (left vs. right hand). BrainRVQ achieves 0.717 AUROC, surpassing CBraMod (0.699) by 2.6\% relative improvement. Motor imagery produces lateralized event-related desynchronization (ERD) patterns in the mu (8--12 Hz) and beta (13--30 Hz) bands over the sensorimotor cortex. The frequency codebook effectively captures these oscillatory signatures, while the hierarchical RVQ structure enables multi-scale representation from coarse motor planning to fine-grained lateralization patterns.

\subsubsection{Motor Imagery Classification (PhysioNet-MI)}

\begin{table}[h]
\centering
\caption{Results on motor imagery classification (PhysioNet-MI, 4-class). Best results are in \textbf{bold}, second best are \underline{underlined}.}
\label{tab:physio_results}
\begin{tabular}{lccc}
\toprule
\textbf{Method} & \textbf{Bal. Acc} & \textbf{Kappa} & \textbf{W-F1} \\
\midrule
EEGNet & 0.581 $\pm$ 0.013 & 0.447 $\pm$ 0.020 & 0.580 $\pm$ 0.012 \\
ST-Transformer & 0.604 $\pm$ 0.008 & 0.471 $\pm$ 0.020 & 0.605 $\pm$ 0.008 \\
BENDR & 0.422 $\pm$ 0.022 & 0.401 $\pm$ 0.025 & 0.408 $\pm$ 0.024 \\
BIOT & 0.615 $\pm$ 0.015 & 0.488 $\pm$ 0.027 & 0.616 $\pm$ 0.020 \\
LaBraM & 0.617 $\pm$ 0.012 & 0.491 $\pm$ 0.019 & 0.618 $\pm$ 0.014 \\
CBraMod & \textbf{0.642 $\pm$ 0.009} & \underline{0.522 $\pm$ 0.017} & \underline{0.643 $\pm$ 0.010} \\
\textbf{\modelname{} (Ours)} & \textbf{0.642 $\pm$ 0.008} & \textbf{0.523 $\pm$ 0.015} & \textbf{0.644 $\pm$ 0.009} \\
\bottomrule
\end{tabular}
\end{table}

Table~\ref{tab:physio_results} presents the results on PhysioNet-MI 4-class motor imagery (left fist, right fist, both fists, both feet). BrainRVQ achieves 0.523 Cohen's Kappa and 0.644 Weighted F1, matching or slightly outperforming CBraMod across all metrics. This task requires distinguishing between unilateral and bilateral motor imagery as well as upper and lower limb movements. The 64-channel high-density recording provides rich spatial information, and our model effectively leverages the pre-trained spatial position embeddings to capture the topographic distribution of motor-related activations across the scalp.

\subsubsection{Emotion Recognition (SEED-V)}

\begin{table}[h]
\centering
\caption{Results on emotion recognition (SEED-V, 5-class). Best results are in \textbf{bold}, second best are \underline{underlined}.}
\label{tab:seedv_results}
\begin{tabular}{lccc}
\toprule
\textbf{Method} & \textbf{Bal. Acc} & \textbf{Kappa} & \textbf{W-F1} \\
\midrule
EEGNet & 0.296 $\pm$ 0.010 & 0.101 $\pm$ 0.014 & 0.275 $\pm$ 0.010 \\
ST-Transformer & 0.305 $\pm$ 0.007 & 0.108 $\pm$ 0.012 & 0.283 $\pm$ 0.011 \\
BENDR & 0.223 $\pm$ 0.006 & 0.034 $\pm$ 0.006 & 0.203 $\pm$ 0.033 \\
BIOT & 0.384 $\pm$ 0.019 & 0.226 $\pm$ 0.026 & 0.386 $\pm$ 0.020 \\
LaBraM & 0.398 $\pm$ 0.014 & 0.239 $\pm$ 0.021 & 0.397 $\pm$ 0.011 \\
CBraMod & \textbf{0.409 $\pm$ 0.010} & \underline{0.257 $\pm$ 0.014} & \underline{0.410 $\pm$ 0.011} \\
\textbf{\modelname{} (Ours)} & \underline{0.407 $\pm$ 0.008} & \textbf{0.260 $\pm$ 0.010} & \textbf{0.414 $\pm$ 0.011} \\
\bottomrule
\end{tabular}
\end{table}

Table~\ref{tab:seedv_results} shows the results on SEED-V 5-class emotion recognition (happy, sad, neutral, disgust, fear). BrainRVQ achieves 0.260 Cohen's Kappa and 0.414 Weighted F1, outperforming CBraMod on both agreement and per-class metrics while achieving comparable balanced accuracy. Emotion recognition from short 1-second EEG segments is particularly challenging due to the transient nature of affective neural responses and high inter-subject variability. The improvement demonstrates that our dual-domain tokenization effectively captures emotion-related patterns, including frontal alpha asymmetry associated with approach/withdrawal motivation and beta/gamma activations related to emotional arousal.

\subsubsection{Abnormal Detection (TUAB)}

\begin{table}[h]
\centering
\caption{Results on abnormal detection (TUAB, 2-class). Best results are in \textbf{bold}, second best are \underline{underlined}.}
\label{tab:tuab_results}
\begin{tabular}{lccc}
\toprule
\textbf{Method} & \textbf{Bal. Acc} & \textbf{AUC-PR} & \textbf{AUROC} \\
\midrule
EEGNet & 0.764 $\pm$ 0.004 & 0.830 $\pm$ 0.004 & 0.841 $\pm$ 0.003 \\
ST-Transformer & 0.797 $\pm$ 0.002 & 0.852 $\pm$ 0.003 & 0.871 $\pm$ 0.002 \\
BENDR & 0.771 $\pm$ 0.025 & 0.841 $\pm$ 0.022 & 0.843 $\pm$ 0.024 \\
BIOT & 0.796 $\pm$ 0.006 & 0.879 $\pm$ 0.002 & 0.882 $\pm$ 0.004 \\
LaBraM & 0.814 $\pm$ 0.002 & 0.897 $\pm$ 0.002 & \underline{0.902 $\pm$ 0.001} \\
CBraMod & \underline{0.829 $\pm$ 0.002} & \textbf{0.926 $\pm$ 0.001} & \textbf{0.923 $\pm$ 0.001} \\
\textbf{\modelname{} (Ours)} & \textbf{0.831 $\pm$ 0.001} & \underline{0.904 $\pm$ 0.001} & 0.900 $\pm$ 0.002 \\
\bottomrule
\end{tabular}
\end{table}

Table~\ref{tab:tuab_results} presents the results on TUAB abnormal detection. BrainRVQ achieves the highest balanced accuracy (0.831), indicating strong performance in correctly classifying both normal and abnormal samples. While CBraMod achieves higher ranking metrics (AUROC 0.923, AUC-PR 0.926), our model demonstrates more balanced predictions across classes. This clinical diagnosis task requires distinguishing normal brain activity from various pathological patterns including slowing, asymmetry, and epileptiform discharges. 

\section{Ablation Study on RVQ Depth}
\label{app:rvq_depth}

To validate the choice of 3-layer Residual Vector Quantization in our DD-RVQ tokenizer, we conduct ablation experiments on downstream tasks with varying numbers of RVQ layers. We evaluate configurations from 1 to 7 layers on Mental Workload and BCICIV-2a datasets. The results are presented in Table~\ref{tab:rvq_depth}.

\begin{table}[h]
\centering
\caption{Ablation study on RVQ depth. Best results are in \textbf{bold}, second best are \underline{underlined}.}
\label{tab:rvq_depth}
\setlength{\tabcolsep}{4pt}
\begin{tabular}{lcccccc}
\toprule
& \multicolumn{3}{c}{Mental Workload, 2-class} & \multicolumn{3}{c}{BCICIV-2a, 4-class} \\
\cmidrule(lr){2-4} \cmidrule(lr){5-7}
RVQ Layers & Bal. Acc & AUC-PR & AUROC & Bal. Acc & Kappa & W-F1 \\
\midrule
1 & 0.740 $\pm$ 0.010 & 0.756 $\pm$ 0.010 & 0.860 $\pm$ 0.012 & 0.491 $\pm$ 0.011 & 0.322 $\pm$ 0.008 & 0.480 $\pm$ 0.013 \\
2 & \underline{0.741 $\pm$ 0.015} & \textbf{0.761 $\pm$ 0.013} & \textbf{0.863 $\pm$ 0.010} & 0.534 $\pm$ 0.015 & \underline{0.383 $\pm$ 0.010} & \textbf{0.551 $\pm$ 0.016} \\
3 (Ours) & \textbf{0.747 $\pm$ 0.011} & \underline{0.758 $\pm$ 0.012} & \underline{0.862 $\pm$ 0.010} & \textbf{0.541 $\pm$ 0.008} & \textbf{0.388 $\pm$ 0.008} & \underline{0.533 $\pm$ 0.012} \\
5 & 0.724 $\pm$ 0.016 & 0.721 $\pm$ 0.018 & 0.816 $\pm$ 0.014 & \underline{0.536 $\pm$ 0.014} & 0.382 $\pm$ 0.013 & 0.528 $\pm$ 0.015 \\
7 & 0.701 $\pm$ 0.012 & 0.692 $\pm$ 0.018 & 0.824 $\pm$ 0.016 & 0.503 $\pm$ 0.012 & 0.337 $\pm$ 0.014 & 0.489 $\pm$ 0.013 \\
\bottomrule
\end{tabular}
\end{table}

The results reveal several important observations. First, single-layer RVQ shows severe degradation on the challenging 4-class motor imagery task (BCICIV-2a), with a 17.0\% drop in Kappa (0.322 vs. 0.388) compared to 3-layer, confirming that multi-layer hierarchical quantization is essential for capturing fine-grained motor-related patterns. Second, 2-layer and 3-layer configurations achieve the best performance across both tasks, with 3-layer achieving the highest balanced accuracy on both datasets (0.747 on Mental Workload and 0.541 on BCICIV-2a) and the best Kappa (0.388) on BCICIV-2a, while 2-layer achieves the best AUC-PR and AUROC on Mental Workload. This suggests that 2-3 layers provide sufficient representational capacity for diverse EEG patterns. Third, deeper configurations (5 and 7 layers) lead to consistent performance degradation, particularly on Mental Workload where 5-layer achieves only 0.816 AUROC compared to 0.862 for 3-layer. This degradation likely results from optimization difficulties with deeper residual structures and potential overfitting to quantization noise.

The task-dependent patterns are also noteworthy. On Mental Workload, a binary classification task, configurations from 1-3 layers achieve similar balanced accuracy (0.740--0.747), suggesting that coarse-grained features suffice for distinguishing stress states. In contrast, BCICIV-2a requires finer discrimination between four motor imagery classes, where the gap between 1-layer (0.322 Kappa) and 3-layer (0.388 Kappa) is substantial. These findings support our choice of 3-layer RVQ as a balanced configuration that achieves strong reconstruction quality while maintaining robust downstream task performance across diverse BCI applications.

\section{Ablation on Mask Ratio}
\label{app:ablation}

\begin{figure}[H]
  \centering
  \includegraphics[width=\textwidth]{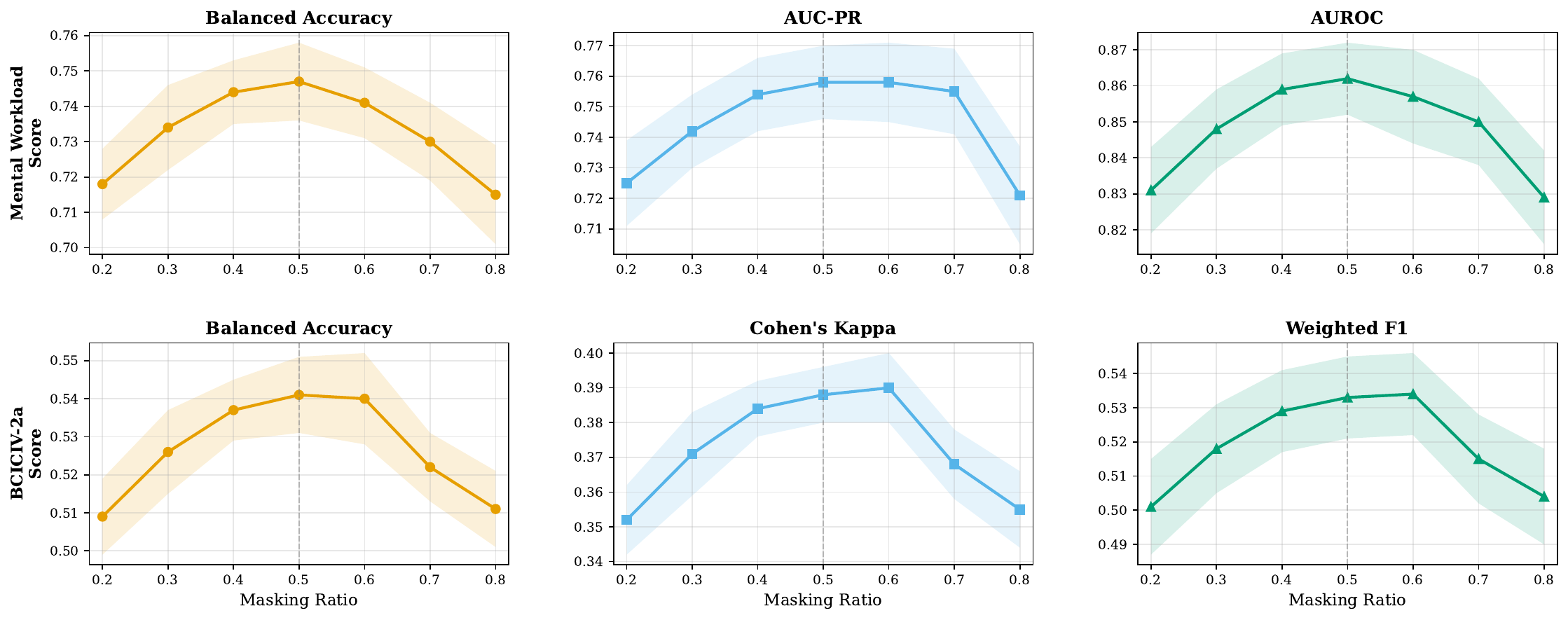}
  \caption{Impact of masking ratio on downstream performance across six metrics. 
  \textbf{Top Row:} Mental Workload dataset metrics (Balanced Accuracy, AUC-PR, AUROC). 
  \textbf{Bottom Row:} BCICIV-2a dataset metrics (Balanced Accuracy, Cohen's Kappa, Weighted F1). 
  Performance consistently peaks at moderate masking ratios (0.5), validating the inverted U-shaped trend. Shaded regions represent standard deviation.}
  \label{fig:mask_ratio_trend}
\end{figure}

We investigate the sensitivity of our pre-training framework to the masking ratio, a critical hyperparameter governing the difficulty of the self-supervised task. Tables~\ref{tab:mask_ratio_mental}, Tables~\ref{tab:mask_ratio_bci} and Figure~\ref{fig:mask_ratio_trend} summarize the downstream performance on the Mental Workload and BCICIV-2a datasets. We observe a consistent inverted U-shaped trend across both tasks, where performance peaks at moderate masking ratios (0.4--0.6) and deteriorates at extremes. Specifically, a masking ratio of 0.5 consistently yields optimal or near-optimal results, achieving the best balance between information redundancy and reconstruction difficulty.

\begin{table}[h]
\centering
\caption{Performance of BrainRVQ on Mental Workload Dataset under Different Mask Ratios.}
\label{tab:mask_ratio_mental}
\setlength{\tabcolsep}{8pt}
\begin{tabular}{lccc}
\toprule
Mask Ratio & Balanced Accuracy & AUC-PR & AUROC \\
\midrule
0.2 & 0.718 $\pm$ 0.010 & 0.725 $\pm$ 0.014 & 0.831 $\pm$ 0.012 \\
0.3 & 0.734 $\pm$ 0.012 & 0.742 $\pm$ 0.012 & 0.848 $\pm$ 0.011 \\
0.4 & \underline{0.744 $\pm$ 0.009} & 0.754 $\pm$ 0.012 & \underline{0.859 $\pm$ 0.010} \\
0.5 & \textbf{0.747 $\pm$ 0.011} & \textbf{0.758 $\pm$ 0.012} & \textbf{0.862 $\pm$ 0.010} \\
0.6 & 0.741 $\pm$ 0.010 & \textbf{0.758 $\pm$ 0.013} & 0.857 $\pm$ 0.013 \\
0.7 & 0.730 $\pm$ 0.011 & 0.755 $\pm$ 0.014 & 0.850 $\pm$ 0.012 \\
0.8 & 0.715 $\pm$ 0.014 & 0.721 $\pm$ 0.016 & 0.829 $\pm$ 0.013 \\
\bottomrule
\end{tabular}
\end{table}

\begin{table}[h]
\centering
\caption{Performance of BrainRVQ on BCICIV-2a Dataset under Different Mask Ratios.}
\label{tab:mask_ratio_bci}
\setlength{\tabcolsep}{8pt}
\begin{tabular}{lccc}
\toprule
Mask Ratio & Balanced Accuracy & Cohen's Kappa & Weighted F1 \\
\midrule
0.2 & 0.509 $\pm$ 0.010 & 0.352 $\pm$ 0.010 & 0.501 $\pm$ 0.014 \\
0.3 & 0.526 $\pm$ 0.011 & 0.371 $\pm$ 0.012 & 0.518 $\pm$ 0.013 \\
0.4 & 0.537 $\pm$ 0.008 & 0.384 $\pm$ 0.008 & 0.529 $\pm$ 0.012 \\
0.5 & \textbf{0.541 $\pm$ 0.010} & \underline{0.388 $\pm$ 0.008} & \underline{0.533 $\pm$ 0.012} \\
0.6 & \underline{0.540 $\pm$ 0.012} & \textbf{0.390 $\pm$ 0.010} & \textbf{0.534 $\pm$ 0.012} \\
0.7 & 0.522 $\pm$ 0.009 & 0.368 $\pm$ 0.010 & 0.515 $\pm$ 0.013 \\
0.8 & 0.511 $\pm$ 0.010 & 0.355 $\pm$ 0.011 & 0.504 $\pm$ 0.014 \\
\bottomrule
\end{tabular}
\end{table}

This phenomenon can be attributed to the trade-off in self-supervised learning dynamics. At low masking ratios (0.2--0.3), the abundance of visible context allows the model to solve the reconstruction task via trivial local interpolation, hindering the learning of semantic global representations. Conversely, excessively high ratios (0.7--0.8) remove too much structural information, turning the prediction task into an ill-posed problem that impedes effective convergence. Notably, the BCICIV-2a dataset exhibits higher sensitivity to masking variations compared to Mental Workload. We hypothesize that this is due to its smaller sample size and the complex nature of motor imagery signals, which necessitate a more precisely tuned pre-training curriculum to extract robust features without overfitting or underfitting.

\section{Visualization}
\label{app:visualization}

This appendix provides visualizations to understand the reconstruction quality of the DD-RVQ tokenizer.

\subsection{Reconstruction Quality Analysis}

\begin{figure}[htb]
    \centering
    \includegraphics[width=\textwidth]{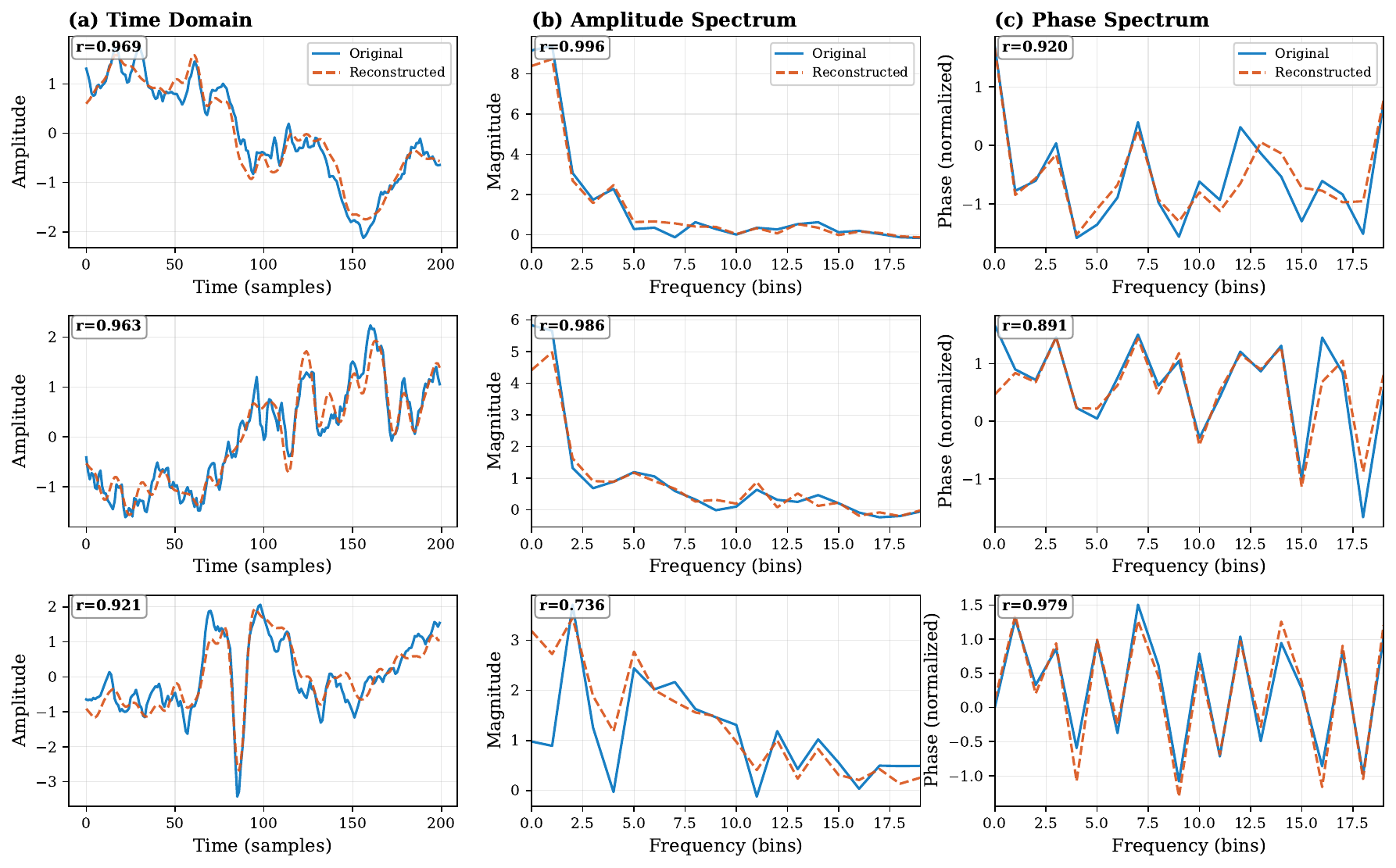}
    \caption{Reconstruction examples from the DD-RVQ tokenizer. Each row shows a different EEG segment. (a) Time-domain waveform reconstruction. (b) Amplitude spectrum reconstruction. (c) Phase spectrum reconstruction. Blue: original signal; Orange dashed: reconstructed signal. Pearson correlation coefficients ($r$) are displayed for each panel.}
    \label{fig:reconstruction_vis}
\end{figure}

We evaluate the reconstruction fidelity of the DD-RVQ tokenizer on a subset of 7,928 EEG samples from the pre-training corpus. Figure~\ref{fig:reconstruction_vis} presents qualitative examples, while Table~\ref{tab:recon_metrics} summarizes quantitative metrics across three reconstruction targets: time-domain waveforms, amplitude spectra, and phase spectra.

The visualized examples demonstrate high reconstruction fidelity across all three domains. In the time domain, the tokenizer accurately captures both the overall waveform envelope and fine-grained oscillatory patterns, achieving correlations of $r=0.969$, $0.963$, and $0.921$ for the three examples. The amplitude spectrum shows excellent preservation of dominant frequency components, particularly in the low-frequency range where most EEG power resides, with correlations reaching $r=0.996$ and $0.986$. Phase reconstruction exhibits more variability ($r=0.891$--$0.979$), yet still captures the overall phase structure faithfully.

\begin{table}[htbp]
\centering
\caption{Reconstruction quality metrics averaged over 7,928 EEG samples.}
\label{tab:recon_metrics}
\begin{tabular}{lccc}
\toprule
\textbf{Metric} & \textbf{Time Domain} & \textbf{Amplitude} & \textbf{Phase} \\
\midrule
MSE ($\times 10^{-3}$) & 178.62 $\pm$ 50.63 & 64.49 $\pm$ 17.51 & 653.32 $\pm$ 49.92 \\
Correlation & 0.904 $\pm$ 0.030 & 0.956 $\pm$ 0.016 & 0.577 $\pm$ 0.042 \\
SNR (dB) & 8.1 $\pm$ 1.2 & 11.2 $\pm$ 1.2 & 2.0 $\pm$ 0.5 \\
\bottomrule
\end{tabular}
\end{table}

Table~\ref{tab:recon_metrics} presents reconstruction metrics averaged over the entire evaluation set. Amplitude spectrum achieves the highest fidelity ($r=0.956$, SNR=11.2 dB), confirming that spectral power distributions are effectively captured by our frequency-domain codebook. Time-domain reconstruction shows strong performance ($r=0.904$, SNR=8.1 dB), demonstrating that the temporal codebook successfully encodes waveform morphology despite the higher complexity of raw signal modeling. Phase reconstruction is more challenging ($r=0.577$, SNR=2.0 dB), consistent with findings in audio processing where phase is inherently harder to model than magnitude~\citep{zeghidour2021soundstream}. The gap between qualitative examples and quantitative averages reflects natural variability in EEG signals: the tokenizer performs well on typical rhythmic patterns while showing reduced fidelity on complex transients. Importantly, perfect reconstruction is not required for downstream tasks---the discrete tokens need only capture sufficient information for the encoder to learn discriminative representations.

\subsection{Effect of RVQ Depth on Reconstruction}

\begin{figure}[htb]
    \centering
    \includegraphics[width=\textwidth]{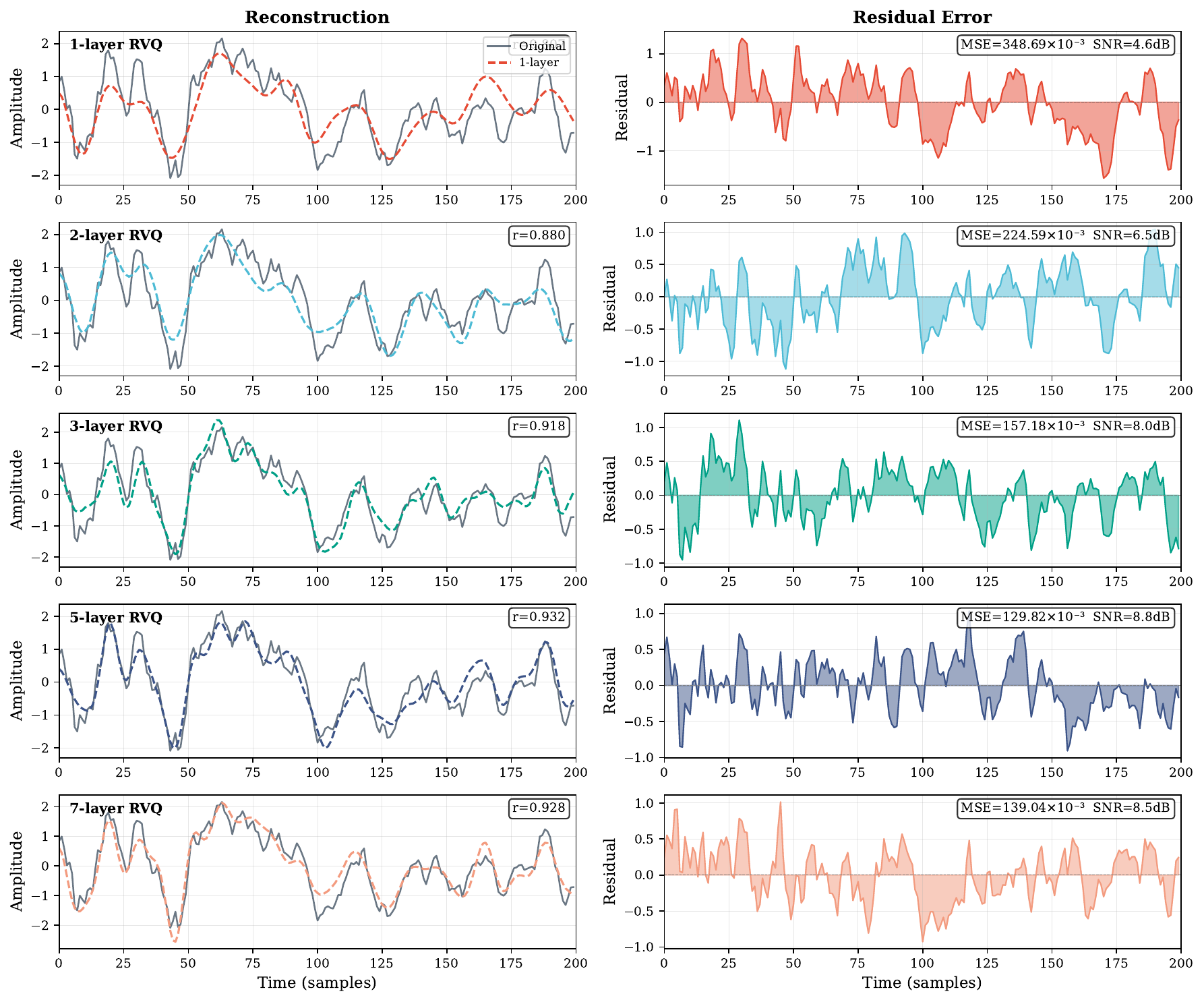}
    \caption{Reconstruction quality with different RVQ depths. Left column: reconstructed waveforms (colored dashed) overlaid on the original signal (gray solid). Right column: residual errors. Correlation coefficients, MSE, and SNR are displayed for each configuration.}
    \label{fig:rvq_depth}
\end{figure}

To analyze how codebook depth affects reconstruction quality, we compare reconstructions using different numbers of RVQ layers (1, 2, 3, 5, and 7 layers). Figure~\ref{fig:rvq_depth} visualizes the progressive refinement on a representative EEG segment, and Table~\ref{tab:rvq_depth} quantifies the reconstruction metrics.

\begin{table}[htbp]
\centering
\caption{Reconstruction quality metrics for different RVQ depths.}
\label{tab:rvq_depth}
\begin{tabular}{lccc}
\toprule
\textbf{RVQ Layers} & \textbf{MSE ($\times 10^{-3}$)} & \textbf{Correlation} & \textbf{SNR (dB)} \\
\midrule
1 layer  & 348.69 & 0.807 & 4.6 \\
2 layers & 224.59 & 0.880 & 6.5 \\
3 layers & 157.18 & 0.918 & 8.0 \\
5 layers & 129.82 & 0.932 & 8.8 \\
7 layers & 139.04 & 0.928 & 8.5 \\
\bottomrule
\end{tabular}
\end{table}

As shown in Figure~\ref{fig:rvq_depth} and Table~\ref{tab:rvq_depth}, increasing the number of RVQ layers from 1 to 3 yields substantial improvements in reconstruction fidelity. The 1-layer configuration captures only coarse structure and dominant low-frequency components, achieving $r=0.807$ with visible distortion in fine details. Adding a second layer reduces MSE by 35.6\% (from 348.69 to 224.59) and improves correlation to 0.880, as it captures residual patterns missed by the first layer. The 3-layer configuration further reduces MSE to 157.18 and achieves $r=0.918$, accurately preserving both slow drifts and fast transients.

However, the benefits of additional layers diminish beyond 3 layers. The 5-layer configuration provides only marginal improvement over 3 layers (correlation: 0.918 $\rightarrow$ 0.932, SNR: 8.0 $\rightarrow$ 8.8 dB), while the 7-layer configuration shows slight performance degradation (correlation drops to 0.928, SNR to 8.5 dB). This saturation and eventual decline suggest that deeper RVQ layers primarily model noise-like residuals rather than meaningful signal structure, potentially leading to overfitting.

Based on these observations, we adopt the 3-layer configuration in our final model, which achieves a 2.2$\times$ reduction in MSE and 3.4 dB improvement in SNR compared to 1-layer, while avoiding the diminishing returns of deeper architectures. This choice is further validated by our ablation study on downstream tasks (Section~\ref{sec:rvq_depth}), where the single-layer variant shows an average performance drop of 5.9\% across four benchmarks, with particularly severe degradation on BCICIV-2a (17.3\% drop in Kappa). The consistent benefit of multi-layer RVQ in both reconstruction quality and downstream task performance confirms that the hierarchical residual structure effectively captures multi-scale EEG patterns essential for discriminative representation learning.

\subsection{Masking Strategy Visualization}

\begin{figure}[htb]
    \centering
    \includegraphics[width=\textwidth]{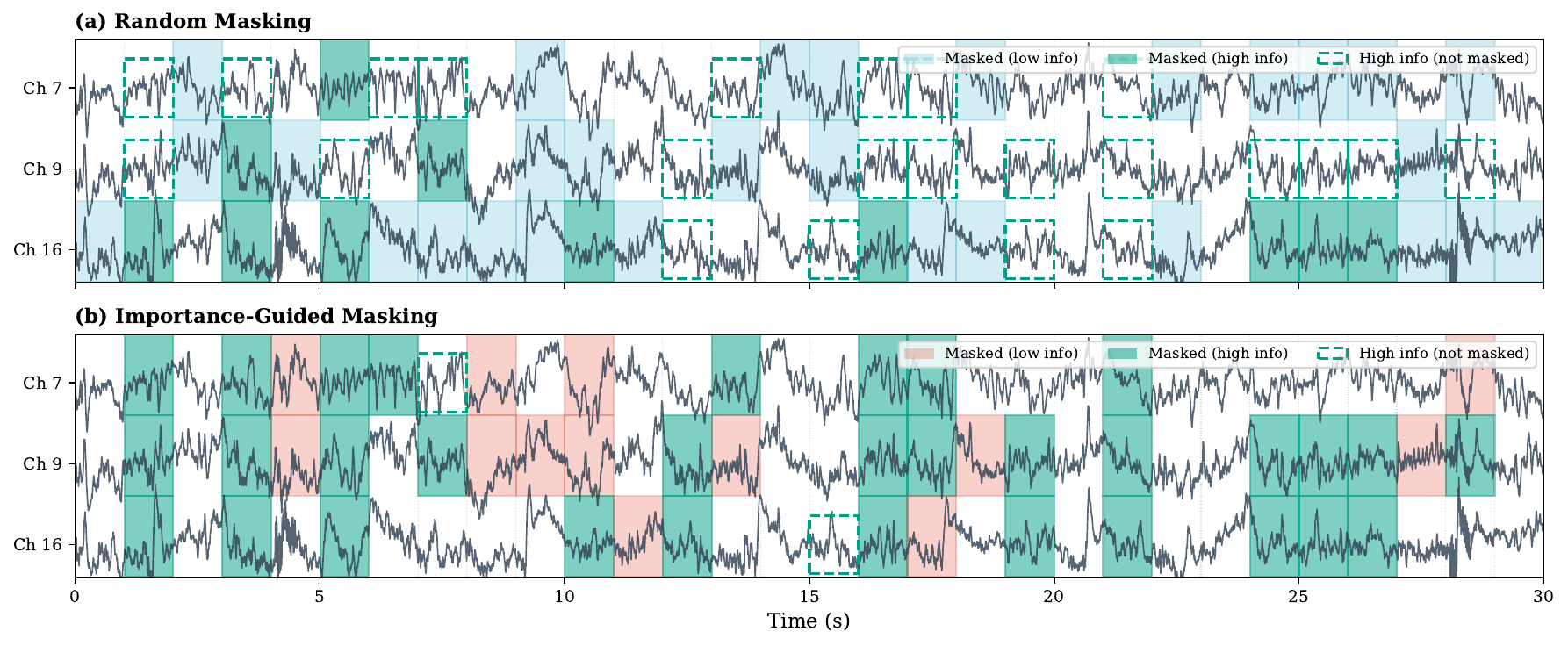}
    \caption{Comparison of mask patterns between random and importance-guided masking. Shaded regions indicate masked patches, with green highlighting high-information patches (top 30\% by information score). Dashed boxes mark high-information patches that remain visible. (a) Random masking distributes masks uniformly, often missing informative regions. (b) Importance-guided masking preferentially targets high-information patches, ensuring the model learns to reconstruct complex neural dynamics.}
    \label{fig:mask_comparison}
\end{figure}

\begin{figure}[htb]
    \centering
    \includegraphics[width=\textwidth]{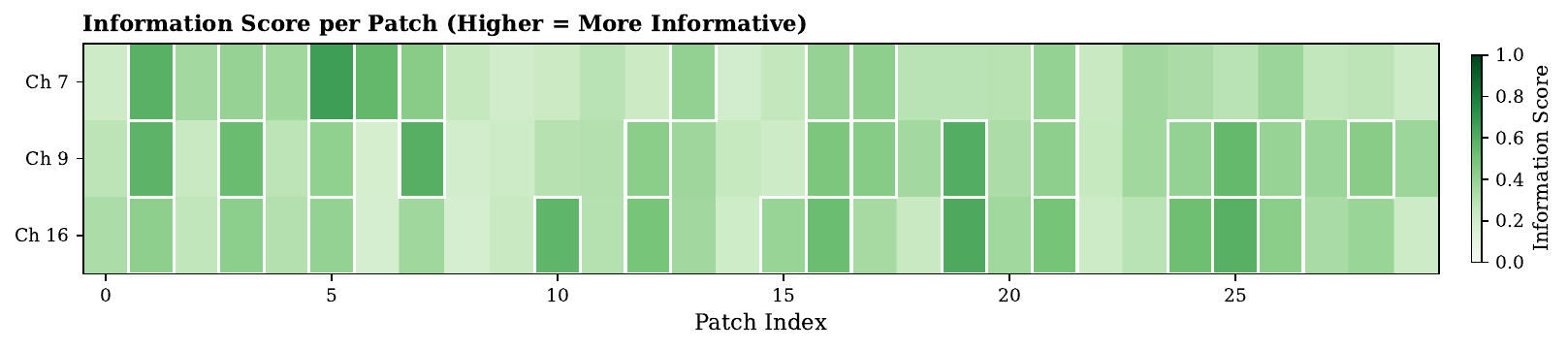}
    \caption{Information score heatmap across patches. Higher scores (darker green) indicate patches with greater neural band power (4--30 Hz), lower artifact contamination, and higher Hjorth complexity. White borders highlight patches in the top 30\% of information scores. The heterogeneous distribution demonstrates that neural information is sparsely distributed across EEG recordings.}
    \label{fig:info_heatmap}
\end{figure}

\begin{figure}[htb]
    \centering
    \includegraphics[width=0.9\textwidth]{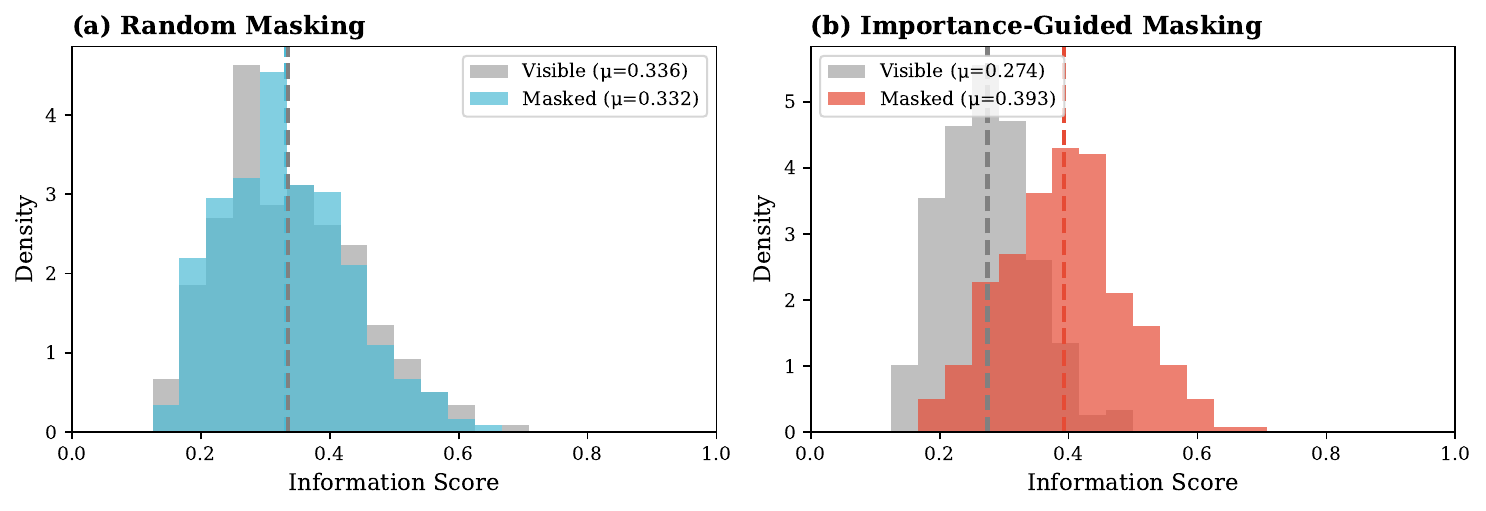}
    \caption{Distribution of information scores for masked vs. visible patches. (a) Under random masking, both distributions are nearly identical. (b) Under importance-guided masking, masked patches have significantly higher mean information scores, confirming that our strategy systematically targets informative regions.}
    \label{fig:mask_distribution}
\end{figure}

To illustrate the difference between random masking and our importance-guided curriculum masking strategy, we visualize the mask patterns generated by both approaches on a representative EEG sample. Figures~\ref{fig:mask_comparison}--\ref{fig:mask_distribution} present comprehensive comparisons from three perspectives: spatial mask patterns, patch-wise information scores, and statistical distributions.

The visualizations reveal a fundamental difference between the two masking strategies. As shown in Figure~\ref{fig:mask_comparison}, random masking distributes masks uniformly across the signal without considering information content, frequently leaving high-information patches (marked with dashed boxes) visible while masking uninformative background regions. In contrast, importance-guided masking systematically captures these informative patches.

The information score heatmap (Figure~\ref{fig:info_heatmap}) further illustrates the heterogeneous and sparse distribution of neural information across EEG recordings. High-scoring regions correspond to segments containing prominent oscillatory activity in the theta-alpha-beta range (4--30 Hz), minimal artifact contamination, and high temporal complexity. By explicitly targeting these informative regions for masking, our curriculum strategy forces the encoder to develop robust representations capable of reconstructing complex neural dynamics from limited context, thereby improving the quality of learned features for downstream classification tasks.

This preferential selection is quantitatively confirmed by Figure~\ref{fig:mask_distribution}: under random masking, masked and visible patches exhibit nearly identical information score distributions ($\mu_{\text{masked}}=0.332$ vs. $\mu_{\text{visible}}=0.336$, $\Delta=-0.004$), whereas importance-guided masking produces a clear separation with masked patches having substantially higher mean scores ($\mu_{\text{masked}}=0.393$ vs. $\mu_{\text{visible}}=0.274$, $\Delta=+0.119$). The information score gap of 0.119 between masked and visible patches under our strategy represents a 43.4\% relative increase compared to the visible set, demonstrating effective prioritization of informative regions.

\section{Scaling Laws Analysis}
\label{app:scaling}

We investigate how BrainRVQ's performance scales with both pre-training data volume and model size, following the scaling law paradigm established in NLP and vision. Understanding these scaling behaviors is crucial for guiding future development of EEG foundation models and determining optimal resource allocation for pre-training. We evaluate on two representative downstream tasks: Mental Workload and BCICIV-2a.

\subsection{Data Scaling}

We pre-train BrainRVQ on varying amounts of data from 1k to 9k hours and evaluate downstream performance. Tables~\ref{tab:data_scaling_mental} and~\ref{tab:data_scaling_bci} summarize the quantitative results.

\begin{table}[H]
\centering
\caption{Data Scaling on Mental Workload Dataset (Binary Classification).}
\label{tab:data_scaling_mental}
\setlength{\tabcolsep}{8pt}
\begin{tabular}{lccc}
\toprule
Data Volume & Balanced Accuracy & AUC-PR & AUROC \\
\midrule
1k hours & 0.695 $\pm$ 0.012 & 0.702 $\pm$ 0.013 & 0.810 $\pm$ 0.010 \\
2k hours & 0.710 $\pm$ 0.013 & 0.718 $\pm$ 0.015 & 0.825 $\pm$ 0.012 \\
3k hours & 0.722 $\pm$ 0.008 & 0.730 $\pm$ 0.014 & 0.838 $\pm$ 0.013 \\
5k hours & 0.732 $\pm$ 0.011 & 0.741 $\pm$ 0.010 & 0.848 $\pm$ 0.005 \\
7k hours & 0.740 $\pm$ 0.010 & 0.750 $\pm$ 0.012 & 0.856 $\pm$ 0.011 \\
\textbf{9k hours} & \textbf{0.747 $\pm$ 0.011} & \textbf{0.758 $\pm$ 0.012} & \textbf{0.862 $\pm$ 0.010} \\
\bottomrule
\end{tabular}
\end{table}

\begin{table}[H]
\centering
\caption{Data Scaling on BCICIV-2a Dataset (4-Class Motor Imagery).}
\label{tab:data_scaling_bci}
\setlength{\tabcolsep}{8pt}
\begin{tabular}{lccc}
\toprule
Data Volume & Balanced Accuracy & Cohen's Kappa & Weighted F1 \\
\midrule
1k hours & 0.485 $\pm$ 0.010 & 0.318 $\pm$ 0.015 & 0.478 $\pm$ 0.016 \\
2k hours & 0.502 $\pm$ 0.010 & 0.338 $\pm$ 0.015 & 0.495 $\pm$ 0.011 \\
3k hours & 0.515 $\pm$ 0.011 & 0.354 $\pm$ 0.020 & 0.508 $\pm$ 0.013 \\
5k hours & 0.525 $\pm$ 0.009 & 0.368 $\pm$ 0.011 & 0.518 $\pm$ 0.008 \\
7k hours & 0.534 $\pm$ 0.009 & 0.379 $\pm$ 0.005 & 0.527 $\pm$ 0.015 \\
\textbf{9k hours} & \textbf{0.541 $\pm$ 0.008} & \textbf{0.388 $\pm$ 0.008} & \textbf{0.533 $\pm$ 0.012} \\
\bottomrule
\end{tabular}
\end{table}

Performance improves consistently with increasing data volume across both tasks, with no clear saturation observed at 9k hours. On Mental Workload, AUROC improves from 0.810 (1k hours) to 0.862 (9k hours), representing a 6.4\% relative improvement. On BCICIV-2a, Cohen's Kappa increases from 0.318 to 0.388, a 22.0\% relative gain. The improvement follows an approximate power-law relationship, with diminishing but still positive returns at larger data scales. Notably, BCICIV-2a shows more pronounced improvement with data scaling, likely because motor imagery classification benefits more from diverse pre-training examples that capture varied neural patterns across subjects. These results suggest that EEG foundation models can continue to benefit from additional pre-training data, and the 9k hours used in our experiments does not represent an upper bound on performance.

\subsection{Model Scaling}

We further investigate how model capacity affects downstream performance by varying the number of Transformer layers in the encoder from 4 to 12, resulting in parameter counts ranging from 1.96M to 5.82M. Tables~\ref{tab:model_scaling_mental} and~\ref{tab:model_scaling_bci} present the results.

\begin{table}[H]
\centering
\caption{Model Scaling on Mental Workload Dataset (Binary Classification).}
\label{tab:model_scaling_mental}
\setlength{\tabcolsep}{8pt}
\begin{tabular}{lcccc}
\toprule
Layers & Parameters & Balanced Accuracy & AUC-PR & AUROC \\
\midrule
4 & 1.96M & 0.720 $\pm$ 0.012 & 0.728 $\pm$ 0.014 & 0.835 $\pm$ 0.013 \\
6 & 2.93M & 0.730 $\pm$ 0.011 & 0.738 $\pm$ 0.010 & 0.845 $\pm$ 0.015 \\
8 & 3.89M & 0.738 $\pm$ 0.015 & 0.747 $\pm$ 0.016 & 0.853 $\pm$ 0.011 \\
10 & 4.86M & 0.743 $\pm$ 0.023 & 0.753 $\pm$ 0.018 & 0.858 $\pm$ 0.018 \\
\textbf{12} & \textbf{5.82M} & \textbf{0.747 $\pm$ 0.011} & \textbf{0.758 $\pm$ 0.012} & \textbf{0.862 $\pm$ 0.010} \\
\bottomrule
\end{tabular}
\end{table}

\begin{table}[H]
\centering
\caption{Model Scaling on BCICIV-2a Dataset (4-Class Motor Imagery).}
\label{tab:model_scaling_bci}
\setlength{\tabcolsep}{8pt}
\begin{tabular}{lcccc}
\toprule
Layers & Parameters & Balanced Accuracy & Cohen's Kappa & Weighted F1 \\
\midrule
4 & 1.96M & 0.508 $\pm$ 0.012 & 0.345 $\pm$ 0.013 & 0.501 $\pm$ 0.014 \\
6 & 2.93M & 0.520 $\pm$ 0.006 & 0.360 $\pm$ 0.008 & 0.513 $\pm$ 0.010 \\
8 & 3.89M & 0.530 $\pm$ 0.015 & 0.372 $\pm$ 0.010 & 0.523 $\pm$ 0.012 \\
10 & 4.86M & 0.536 $\pm$ 0.015 & 0.381 $\pm$ 0.012 & 0.529 $\pm$ 0.015 \\
\textbf{12} & \textbf{5.82M} & \textbf{0.541 $\pm$ 0.008} & \textbf{0.388 $\pm$ 0.008} & \textbf{0.533 $\pm$ 0.012} \\
\bottomrule
\end{tabular}
\end{table}

Across both datasets and all metrics, we observe consistent performance gains as model size increases. On Mental Workload, AUROC improves from 0.835 (4 layers, 1.96M) to 0.862 (12 layers, 5.82M), a 3.2\% relative improvement. On BCICIV-2a, Cohen's Kappa increases from 0.345 to 0.388, representing a 12.5\% relative gain. The improvements are particularly pronounced when scaling from 4 to 8 layers, with diminishing returns observed beyond 10 layers. This pattern suggests that while increased model capacity generally benefits representation learning, the marginal gains decrease as the model approaches sufficient expressiveness for the pre-training task.

BCICIV-2a shows stronger sensitivity to model scaling compared to Mental Workload, mirroring the pattern observed in data scaling. This may reflect the higher complexity of 4-class motor imagery classification, which requires finer-grained discriminative features that benefit more from increased model capacity. It is worth noting that our model remains relatively compact (5.82M parameters for the full 12-layer configuration) compared to foundation models in NLP and vision, yet still demonstrates clear scaling benefits.

\section{Low-Resource Fine-tuning Evaluation}
\label{app:low_resource}

In practical BCI applications, obtaining labeled EEG data is often expensive and time-consuming, requiring expert annotation or controlled experimental paradigms. This motivates the need for foundation models that can achieve strong performance with limited labeled data. We evaluate BrainRVQ under low-resource fine-tuning scenarios using only 30\% of available training data.

Table~\ref{tab:low_resource} presents the low-resource fine-tuning results on SEED-V and SHU-MI datasets, comparing BrainRVQ against existing foundation models under both 30\% and 100\% data settings. Bold indicates the best and underline indicates the second best.

\begin{table}[h]
\centering
\caption{Performance comparison on low-resource settings with 30\% of fine-tuning data.}
\label{tab:low_resource}
\setlength{\tabcolsep}{4pt}
\begin{tabular}{lcccccc}
\toprule
& \multicolumn{3}{c}{SEED-V, 5-class} & \multicolumn{3}{c}{SHU-MI, 2-class} \\
\cmidrule(lr){2-4} \cmidrule(lr){5-7}
Methods & Bal. Acc & Kappa & W-F1 & Bal. Acc & AUC-PR & AUROC \\
\midrule
CBraMod (full data) & 0.409 $\pm$ 0.010 & 0.257 $\pm$ 0.014 & 0.410 $\pm$ 0.011 & \textbf{0.637 $\pm$ 0.015} & 0.714 $\pm$ 0.009 & 0.699 $\pm$ 0.007 \\
\textbf{Ours} (full data) & \textbf{0.407 $\pm$ 0.008} & \textbf{0.260 $\pm$ 0.010} & \textbf{0.414 $\pm$ 0.011} & \textbf{0.637 $\pm$ 0.018} & \textbf{0.723 $\pm$ 0.011} & \textbf{0.717 $\pm$ 0.007} \\
\midrule
BIOT (30\%) & 0.351 $\pm$ 0.038 & 0.178 $\pm$ 0.043 & 0.349 $\pm$ 0.042 & 0.542 $\pm$ 0.035 & 0.626 $\pm$ 0.027 & 0.602 $\pm$ 0.030 \\
LaBraM (30\%) & 0.369 $\pm$ 0.031 & 0.204 $\pm$ 0.038 & 0.370 $\pm$ 0.032 & 0.564 $\pm$ 0.029 & 0.652 $\pm$ 0.023 & 0.647 $\pm$ 0.027 \\
CBraMod (30\%) & \underline{0.388 $\pm$ 0.024} & \underline{0.229 $\pm$ 0.025} & \underline{0.389 $\pm$ 0.026} & \textbf{0.623 $\pm$ 0.020} & \underline{0.690 $\pm$ 0.013} & \underline{0.675 $\pm$ 0.011} \\
\textbf{Ours} (30\%) & \textbf{0.390 $\pm$ 0.030} & \textbf{0.238 $\pm$ 0.032} & \textbf{0.396 $\pm$ 0.030} & \underline{0.622 $\pm$ 0.018} & \textbf{0.692 $\pm$ 0.022} & \textbf{0.685 $\pm$ 0.018} \\
\bottomrule
\end{tabular}
\end{table}

The results demonstrate that BrainRVQ achieves competitive or superior performance compared to existing foundation models in low-resource settings. On SEED-V, BrainRVQ with 30\% data achieves 0.238 Cohen's Kappa, outperforming CBraMod (0.229) and LaBraM (0.204) by 3.9\% and 16.7\% respectively. On SHU-MI, our 30\% result achieves the best AUROC (0.685) and AUC-PR (0.692), surpassing CBraMod's 30\% performance (AUROC=0.675, AUC-PR=0.690), while maintaining comparable balanced accuracy.

Notably, BrainRVQ with only 30\% of training data on SHU-MI (AUROC=0.685) retains 95.5\% of its full-data performance (AUROC=0.717), demonstrating strong data efficiency. Similarly, on SEED-V, our 30\% result (Kappa=0.238) achieves 91.5\% of the full-data performance (Kappa=0.260). These observations underscore the practical utility of our approach for data-scarce BCI applications where annotation costs are prohibitive.

\section{Discussion}
\label{app:broader_impacts}

\subsection{Broader Impacts}

The development of EEG foundation models such as BrainRVQ represents an important step forward in the intersection of AI and neurophysiological signal processing. We discuss several potential broader impacts of our work.

\textbf{Clinical Applications.} BrainRVQ has significant potential for clinical EEG analysis. For seizure detection, automated monitoring of epilepsy patients could substantially reduce clinician workload and enable faster intervention during ictal events. For sleep disorders, accurate automated sleep staging supports diagnosis of conditions such as sleep apnea, insomnia, and narcolepsy, where manual scoring by trained technicians is time-consuming and subject to inter-rater variability. For neurological screening, abnormality detection capabilities could assist in triaging patients who require further neurological evaluation, potentially enabling earlier diagnosis of conditions such as encephalopathy or focal brain lesions.

\textbf{Brain-Computer Interface Development.} Our model advances the development of practical BCI systems in several ways. By leveraging large-scale pre-training with dual-domain tokenization, BrainRVQ learns representations that capture both temporal dynamics and spectral characteristics, reducing the need for extensive per-user calibration. The model's demonstrated cross-task generalization supports development of universal BCIs that can adapt to multiple applications without task-specific retraining. Improved motor imagery and emotion recognition decoding could enhance assistive communication devices for patients with severe motor disabilities such as amyotrophic lateral sclerosis (ALS) or locked-in syndrome.

\textbf{Scientific Understanding.} Beyond practical applications, the learned representations provide insights into neural dynamics. The hierarchical discrete token representation enables analysis of which neural patterns are most informative at different granularities, from coarse signal structure to fine-grained details. The dual-domain codebooks offer interpretable decomposition of EEG signals into temporal waveform patterns and frequency-domain characteristics, potentially revealing how these complementary aspects contribute to different cognitive and clinical tasks.

\subsection{Limitations}

Our work has several limitations. First, the model is pre-trained on TUEG from a single medical center, which may limit generalization to EEG recorded with different equipment, protocols, or patient populations. Future work should curate diverse multi-center pre-training corpora to reduce institutional bias and improve cross-site generalization. Second, pre-training uses 19 channels conforming to the 10-20 system. While the model transfers to other channel configurations through flexible positional encoding, optimal performance may require similar electrode placements. Developing channel-agnostic architectures that seamlessly adapt to arbitrary montages remains an important challenge. Third, our current framework focuses solely on EEG. Incorporating other physiological signals such as EOG, EMG, and ECG could enhance performance on tasks like sleep staging where multi-modal information is clinically relevant.

\subsection{Future Work}

Looking forward, we envision several promising research directions. First, scaling pre-training to larger models with more diverse EEG data from multiple institutions could further improve generalization. Second, exploring more sophisticated masking strategies beyond importance-guided curriculum masking may enhance representation learning. Third, extending the dual-domain framework to incorporate additional signal domains (e.g., time-frequency representations) could capture richer neural patterns. Finally, conducting prospective clinical validation studies and developing continual learning methods for efficient model updating would facilitate real-world deployment.

\end{document}